\documentclass[%
reprint,
superscriptaddress,
showpacs,preprintnumbers,
amsmath,amssymb,
aps,
floatfix,
]{revtex4-1}

\usepackage{graphicx}
\usepackage{dcolumn}
\usepackage{bm}
\usepackage[section]{placeins}

\begin{document}

\preprint{APS/123-QED}

\title{Helicity-dependent cross sections and double-polarization observable \boldmath{$E$} in 
$\bm{\eta}$ photoproduction from quasi-free protons and neutrons}

\author{L.~Witthauer}
\affiliation{Department of Physics, University of Basel, CH-4056 Basel, Switzerland}
\author {M.~Dieterle}
\affiliation{Department of Physics, University of Basel, CH-4056 Basel, Switzerland}
\author{S.~Abt}
\affiliation{Department of Physics, University of Basel, CH-4056 Basel, Switzerland}
\author{P.~Achenbach}
\affiliation{Institut f\"ur Kernphysik, University of Mainz, D-55099 Mainz, Germany}
\author{F.~Afzal}
\affiliation{Helmholtz-Institut f\"ur Strahlen- und Kernphysik, University of Bonn, D-53115 Bonn, Germany}
\author{Z.~Ahmed}
\affiliation{University of Regina, Regina, SK S4S 0A2 Canada}
\author {C.S.~Akondi}
\affiliation{Kent State University, Kent, Ohio 44242-0001, USA}
\author{J.R.M.~Annand}
\affiliation{SUPA School of Physics and Astronomy, University of Glasgow, Glasgow, G12 8QQ, UK}
\author{H.J.~Arends}
\affiliation{Institut f\"ur Kernphysik, University of Mainz, D-55099 Mainz, Germany}
\author{M.~Bashkanov}
\affiliation{SUPA School of Physics, University of Edinburgh, Edinburgh EEH9 3JZ, UK}
\author{R.~Beck}
\affiliation{Helmholtz-Institut f\"ur Strahlen- und Kernphysik, University of Bonn, D-53115 Bonn, Germany}
\author{M.~Biroth}
\affiliation{Institut f\"ur Kernphysik, University of Mainz, D-55099 Mainz, Germany}
\author{N.S.~Borisov}
\affiliation{Joint Institute for Nuclear Research, 141980 Dubna, Russia}
\author{A.~Braghieri}
\affiliation{INFN Sezione di Pavia, I-27100 Pavia, Italy}
\author{W.J.~Briscoe}
\affiliation{Center for Nuclear Studies, The George Washington University, Washington, DC 20052-0001, USA}
\author{ F.~Cividini}
\affiliation{Institut f\"ur Kernphysik, University of Mainz, D-55099 Mainz, Germany}
\author{S.~Costanza}\altaffiliation{Also at Dipartimento di Fisica, Universit\`a di Pavia, I-27100 Pavia, Italy.}
\affiliation{INFN Sezione di Pavia, I-27100 Pavia, Italy}
\author{C.~Collicott}
\affiliation{Department of Astronomy and Physics, Saint Marys University, Halifax, Nova Scotia B3H 3C3, Canada}
\author{A.~Denig}
\affiliation{Institut f\"ur Kernphysik, University of Mainz, D-55099 Mainz, Germany}
\author{E.J.~Downie}
\affiliation{Institut f\"ur Kernphysik, University of Mainz, D-55099 Mainz, Germany}
\affiliation{Center for Nuclear Studies, The George Washington University, Washington, DC 20052-0001, USA}
\author{P.~Drexler}
\affiliation{Institut f\"ur Kernphysik, University of Mainz, D-55099 Mainz, Germany}
\author{M.I.~Ferretti-Bondy}
\affiliation{Institut f\"ur Kernphysik, University of Mainz, D-55099 Mainz, Germany}
\author{S.~Gardner}
\affiliation{SUPA School of Physics and Astronomy, University of Glasgow, Glasgow, G12 8QQ, UK}
\author{S.~Garni}
\affiliation{Department of Physics, University of Basel, CH-4056 Basel, Switzerland}
\author{D.I.~Glazier}
\affiliation{SUPA School of Physics and Astronomy, University of Glasgow, Glasgow, G12 8QQ, UK}
\affiliation{SUPA School of Physics, University of Edinburgh, Edinburgh EEH9 3JZ, UK}
\author{D.~Glowa}
\affiliation{SUPA School of Physics, University of Edinburgh, Edinburgh EEH9 3JZ, UK}
\author{W.~Gradl}
\affiliation{Institut f\"ur Kernphysik, University of Mainz, D-55099 Mainz, Germany}
\author{M.~G\"unther}
\affiliation{Department of Physics, University of Basel, CH-4056 Basel, Switzerland}
\author{G.M.~Gurevich}
\affiliation{Institute for Nuclear Research, 125047 Moscow, Russia}
\author{D.~Hamilton}
\affiliation{SUPA School of Physics and Astronomy, University of Glasgow, Glasgow, G12 8QQ, UK}
\author{D.~Hornidge}
\affiliation{Mount Allison University, Sackville, New Brunswick E4L 1E6, Canada}
\author{G.M.~Huber}
\affiliation{University of Regina, Regina, SK S4S 0A2 Canada}
\author{A.~K{\"a}ser}
\affiliation{Department of Physics, University of Basel, CH-4056 Basel, Switzerland}
\author{V.L.~Kashevarov}
\affiliation{Institut f\"ur Kernphysik, University of Mainz, D-55099 Mainz, Germany}
\author{ S.~Kay}
\affiliation{SUPA School of Physics, University of Edinburgh, Edinburgh EEH9 3JZ, UK}
\author{I.~Keshelashvili}\altaffiliation{Now at Institut f\"ur Kernphysik, FZ J\"ulich, 52425 J\"ulich, Germany}
\affiliation{Department of Physics, University of Basel, CH-4056 Basel, Switzerland}
\author{R.~Kondratiev}
\affiliation{Institute for Nuclear Research, 125047 Moscow, Russia}
\author{M.~Korolija}
\affiliation{Rudjer Boskovic Institute, HR 10000 Zagreb, Croatia}
\author{B.~Krusche}\email[]{Corresponding author: email bernd.krusche@unibas.ch}
\affiliation{Department of Physics, University of Basel, CH-4056 Basel, Switzerland}
\author{A.B.~Lazarev}
\affiliation{Joint Institute for Nuclear Research, 141980 Dubna, Russia}
\author{J.M.~Linturi}
\affiliation{Institut f\"ur Kernphysik, University of Mainz, D-55099 Mainz, Germany}
\author{V.~Lisin}
\affiliation{Institute for Nuclear Research, 125047 Moscow, Russia}
\author{K.~Livingston}
\affiliation{SUPA School of Physics and Astronomy, University of Glasgow, Glasgow, G12 8QQ, UK}
\author{S.~Lutterer}
\affiliation{Department of Physics, University of Basel, CH-4056 Basel, Switzerland}
\author{I.J.D.~MacGregor}
\affiliation{SUPA School of Physics and Astronomy, University of Glasgow, Glasgow, G12 8QQ, UK}
\author{J.~Mancell}
\affiliation{SUPA School of Physics and Astronomy, University of Glasgow, Glasgow, G12 8QQ, UK}
\author{D.M.~Manley}
\affiliation{Kent State University, Kent, Ohio 44242-0001, USA}
\author{P.P.~Martel}
\affiliation{Institut f\"ur Kernphysik, University of Mainz, D-55099 Mainz, Germany}
\affiliation{Mount Allison University, Sackville, New Brunswick E4L 1E6, Canada}
\author{V.~Metag}
\affiliation{II. Physikalisches Institut, University of Giessen, D-35392 Giessen, Germany}
\author{W.~Meyer}
\affiliation{Institut f\"ur Experimentalphysik, Ruhr Universit\"at, 44780 Bochum, Germany}
\author{R.~Miskimen}
\affiliation{University of Massachusetts Amherst, Amherst, Massachusetts 01003, USA}
\author{E.~Mornacchi}
\affiliation{Institut f\"ur Kernphysik, University of Mainz, D-55099 Mainz, Germany}
\author{A.~Mushkarenkov}
\affiliation{Institute for Nuclear Research, 125047 Moscow, Russia}
\affiliation{University of Massachusetts Amherst, Amherst, Massachusetts 01003, USA}
\author{A.B.~Neganov}
\affiliation{Joint Institute for Nuclear Research, 141980 Dubna, Russia}
\author{A.~Neiser}
\affiliation{Institut f\"ur Kernphysik, University of Mainz, D-55099 Mainz, Germany}
\author{M.~Oberle}
\affiliation{Department of Physics, University of Basel, CH-4056 Basel, Switzerland}
\author{M.~Ostrick} 
\affiliation{Institut f\"ur Kernphysik, University of Mainz, D-55099 Mainz, Germany}
\author{P.B.~Otte}
\affiliation{Institut f\"ur Kernphysik, University of Mainz, D-55099 Mainz, Germany}
\author{ D.~Paudyal}
\affiliation{University of Regina, Regina, SK S4S 0A2 Canada}
\author{P.~Pedroni}
\affiliation{INFN Sezione di Pavia, I-27100 Pavia, Italy}
\author{A.~Polonski}
\affiliation{Institute for Nuclear Research, 125047 Moscow, Russia}
\author{S.N.~Prakhov}
\affiliation{Institut f\"ur Kernphysik, University of Mainz, D-55099 Mainz, Germany}
\affiliation{University of California at Los Angeles, Los Angeles, California 90095-1547, USA}
\author{A.~Rajabi}
\affiliation{University of Massachusetts Amherst, Amherst, Massachusetts 01003, USA}
\author{G.~Reicherz}
\affiliation{Institut f\"ur Experimentalphysik, Ruhr Universit\"at, 44780 Bochum, Germany}
\author{G.~Ron}
\affiliation{Racah Institute of Physics, Hebrew University of Jerusalem, Jerusalem 91904, Israel}
\author{T.~Rostomyan}\altaffiliation{Now at Department of Physics and Astronomy., Rutgers University,
Piscataway, New Jersey, 08854-8019, USA}
\affiliation{Department of Physics, University of Basel, CH-4056 Basel, Switzerland}
\author{A.~Sarty}
\affiliation{Department of Astronomy and Physics, Saint Marys University, Halifax, Nova Scotia B3H 3C3, Canada}
\author{C.~Sfienti}
\affiliation{Institut f\"ur Kernphysik, University of Mainz, D-55099 Mainz, Germany}
\author{M.H.~Sikora}
\affiliation{SUPA School of Physics, University of Edinburgh, Edinburgh EEH9 3JZ, UK}
\author{V.~Sokhoyan}
\affiliation{Institut f\"ur Kernphysik, University of Mainz, D-55099 Mainz, Germany}
\affiliation{Center for Nuclear Studies, The George Washington University, Washington, DC 20052-0001, USA}
\author{K.~Spieker}
\affiliation{Helmholtz-Institut f\"ur Strahlen- und Kernphysik, University of Bonn, D-53115 Bonn, Germany}
\author{O.~Steffen}
\affiliation{Institut f\"ur Kernphysik, University of Mainz, D-55099 Mainz, Germany}
\author{I.I.~Strakovsky}
\affiliation{Center for Nuclear Studies, The George Washington University, Washington, DC 20052-0001, USA}
\author{Th.~Strub}
\affiliation{Department of Physics, University of Basel, CH-4056 Basel, Switzerland}
\author{I.~Supek}
\affiliation{Rudjer Boskovic Institute, HR 10000 Zagreb, Croatia}
\author{A.~Thiel}
\affiliation{Helmholtz-Institut f\"ur Strahlen- und Kernphysik, University of Bonn, D-53115 Bonn, Germany}
\author{M.~Thiel}
\affiliation{Institut f\"ur Kernphysik, University of Mainz, D-55099 Mainz, Germany}
\author{A.~Thomas}
\affiliation{Institut f\"ur Kernphysik, University of Mainz, D-55099 Mainz, Germany}
\author{M.~Unverzagt}
\affiliation{Institut f\"ur Kernphysik, University of Mainz, D-55099 Mainz, Germany}
\author{Yu.A.~Usov}
\affiliation{Joint Institute for Nuclear Research, 141980 Dubna, Russia}
\author{S.~Wagner}
\affiliation{Institut f\"ur Kernphysik, University of Mainz, D-55099 Mainz, Germany}
\author{N.K.~Walford}
\affiliation{Department of Physics, University of Basel, CH-4056 Basel, Switzerland}
\author{D.P.~Watts}
\affiliation{SUPA School of Physics, University of Edinburgh, Edinburgh EEH9 3JZ, UK}
\author{D.~Werthm\"uller}
\affiliation{Department of Physics, University of Basel, CH-4056 Basel, Switzerland}
\affiliation{SUPA School of Physics and Astronomy, University of Glasgow, Glasgow, G12 8QQ, UK}
\author{J.~Wettig}
\affiliation{Institut f\"ur Kernphysik, University of Mainz, D-55099 Mainz, Germany}
\author{M.~Wolfes}
\affiliation{Institut f\"ur Kernphysik, University of Mainz, D-55099 Mainz, Germany}
\author{L.~Zana}
\affiliation{SUPA School of Physics, University of Edinburgh, Edinburgh EEH9 3JZ, UK}
\collaboration{A2 Collaboration at MAMI}

\date{\today}

\begin{abstract}
Precise helicity-dependent cross sections and the double-polarization observable $E$ were measured 
for $\eta$ photoproduction from quasi-free protons and neutrons bound in the deuteron. The 
$\eta\rightarrow 2\gamma$ and $\eta\rightarrow 3\pi^0\rightarrow 6\gamma$ decay modes were used to 
optimize the statistical quality of the data and to estimate systematic uncertainties. The measurement
used the A2 detector setup at the tagged photon beam of the electron accelerator MAMI in Mainz. 
A longitudinally polarized deuterated butanol target was used in combination with a circularly
polarized photon beam from bremsstrahlung of a longitudinally polarized electron beam.
The reaction products were detected with the electromagnetic calorimeters Crystal Ball and TAPS, 
which covered 98\% of the full solid angle. The results show that the narrow structure observed earlier
in the unpolarized excitation function of $\eta$ photoproduction off the neutron appears only in reactions 
with antiparallel photon and nucleon spin ($\sigma_{1/2}$). It is absent for reactions with parallel
spin orientation ($\sigma_{3/2}$) and thus very probably related to partial waves with total spin 1/2.
The behavior of the angular distributions of the helicity-dependent cross sections was analyzed by
fitting them with Legendre polynomials. The results are in good agreement with a model from the 
Bonn-Gatchina group, which uses an interference of $P_{11}$ and $S_{11}$ partial waves to explain the
narrow structure. 
\end{abstract}

\maketitle
\section{Introduction}
\label{intro}
During the last few years photoproduction of mesons has been the major source of new experimental
information about nucleon resonances and its impact becomes apparent in the Review of Particle
Physics (RPP) \cite{PDG_14,PDG_16}. This progress has two main roots. The measurement of many different
observables, using polarized beams and polarized targets, allows almost
model independent reaction analyses. A nice example for the progress of the interpretation of
pion production data is given in \cite{Anisovich_16}. The other root is the measurement of many
different final states, which allows coupled channel analyses. Some nucleon - meson final states 
are selective for specific subclasses of nucleon resonances. One of them is photoproduction of
$\eta$ mesons for which (like for $\eta '$ mesons) the selectivity is twofold. Due to the isoscalar 
nature of these mesons only $I = 1/2$ $N^{\star}$ nucleon resonances can decay directly
to the nucleon ground state by their emission. Decays of $\Delta^{\star}$ resonances are possible 
to the $\Delta(1232)$, but this results in $\eta\pi N$ final states, which have recently also been
under detailed investigation \cite{Kaeser_15,Kaeser_16}. Furthermore, due to the relatively large 
masses of these mesons, partial waves with low momenta are preferred even for relatively large 
incident photon energies, making them ideal tools for the search of low-momentum missing resonances
at higher excitation energies. A recent overview of the production of $\eta$, $\eta '$, and $\eta\pi$
pairs is given in \cite{Krusche_15}.        

Photoproduction of $\eta$ mesons off protons has been studied in much detail during the last decade.
A special feature of this reaction is that the kinematic production threshold ($W=1485$~MeV) lies 
just below the Breit-Wigner mass ($W=1535$~MeV) of the s-wave resonance $N(1535) 1/2^{-}$ with a width
of $\approx150$~MeV and a very strong coupling to the $N\eta$ final state (branching ratio 
$b_{\eta}\approx40$\%, the deeper reasons for this strong coupling are not well understood). Therefore, 
photoproduction of $\eta$ mesons in the threshold range is completely dominated by this resonance
\cite{Krusche_95,Krusche_97,Krusche_03}. Other resonances ($N(1520)3/2^{-}$) contribute at threshold 
only via interference terms with the leading $E_{0+}$ multipole \cite{Krusche_03} or at higher
excitation energies \cite{Krusche_15}. Precise measurements of differential cross sections 
have been reported from CLAS \cite{Dugger_02,Williams_09}, ELSA \cite{Crede_05,Bartholomy_07,Crede_09},
GRAAL \cite{Renard_02}, and MAMI \cite{Krusche_95,McNicoll_10}. The beam asymmetry $\Sigma$
has been measured at GRAAL and at ELSA \cite{Ajaka_98,Elsner_07,Bartalini_07}, results for
the target asymmetry $T$ and the double-polarization observable $F$ have been published from
the Crystal Ball/TAPS experiment at MAMI \cite{Akondi_14}, results for the double-polarization 
observable $E$ have been reported from the CLAS experiment \cite{Senderovich_16}, and new results 
for the polarization observables $T$, $E$, $P$, $H$, and $G$ from ELSA have been submitted for 
publication \cite{Mueller_16}. These data will certainly help to identify contributions
from resonances that couple only weakly to $N\eta$. 

The database for photoproduction of $\eta$ mesons off (quasi-free) neutrons $\gamma n\rightarrow n\eta$
is still much less complete, but the study of this reaction is imperative for the isospin decomposition
of the amplitudes. Experiments and also the interpretation of the results for a quasi-free reaction
off nucleons bound in light nuclei like the deuteron are in several aspects more complicated than
measurements of reactions with free proton targets. The necessary detection of the recoil neutrons
in coincidence with the $\eta$-mesons reduces strongly the overall detection efficiency and introduces
additional systematic uncertainties. Typical neutron detection efficiencies in electromagnetic calorimeters 
are on the order of 30\% or less meaning that the detected reaction rates are reduced by
approximately a factor of three compared to measurements with free nucleon targets not requiring
detection of recoil nucleons. All structures in excitation functions are smeared by nuclear Fermi 
motion; furthermore, nuclear final-state interaction (FSI) effects may introduce further complications. 

The unexpected results reported during the last few years for photoproduction of $\eta$ mesons
off neutrons have raised a lot of interest. It came as a surprise when first measurements of the 
$\gamma n\rightarrow n\eta$ excitation function using deuterium targets at the GRAAL facility in Grenoble
\cite{Kuznetsov_07}, at the ELSA accelerator in Bonn \cite{Jaegle_08,Jaegle_11}, and at LNS (now ELPH) in Tohoku
\cite{Miyahara_07} reported a pronounced, very narrow, peak-like structure at nucleon - $\eta$ invariant 
masses close to 1.68~GeV (incident photon energies around 1~GeV). In the meantime, high statistics 
measurements at the MAMI accelerator in Mainz \cite{Werthmueller_13,Witthauer_13,Werthmueller_14}
have established this structure beyond any doubts and investigated in detail its energy dependence and 
angular dependence. Such a structure was not observed for the proton target, although the excitation
function of $\gamma p\rightarrow p\eta$ \cite{McNicoll_10} shows a narrow dip-like structure at the 
same energy. It did not seem unlikely that both structures are related, but so far there is no evidence 
for this and the present results (see Sec.~\ref{sec:Res}) do not favor this conjecture. 

The nature of the narrow structure in the $\gamma n\rightarrow n\eta$ neutron excitation has been 
discussed by several authors in quite different scenarios. Some analyses 
(e.g. Refs.~\cite{Polyakov_03,Arndt_04,Choi_06,Fix_07,Shrestha_12}) interpret it as a new, narrow nucleon 
resonance with partly exotic properties. In the 2014 edition of the RPP \cite{PDG_14} it was listed as
a tentative (one star rating) $N(1685)$ state with otherwise unknown properties, in the 2016 edition it 
was removed again. Other tentative 
explanations include contributions from intermediate strangeness loops \cite{Doering_10} or coupled-channel 
and interference effects of known nucleon resonances \cite{Shklyar_07,Shyam_08}. In the
Bonn-Gatchina (BnGa) coupled-channel analysis a solution was proposed \cite{Anisovich_15} that is based 
on interferences between contributions from the $N(1535)$ and $N(1650)$ spin $1/2$ resonances and 
non-resonant background in the same partial wave. 

Recent experimental developments have further added to this puzzle. Kuznetsov and collaborators 
\cite{Kuznetsov_15} reported results from the GRAAL experiment for the beam asymmetry $\Sigma$ in Compton 
scattering off the free proton, which show a narrow peak at the same energy as the peak in the excitation 
function for $\eta$ production off the neutron. Furthermore, they observed a second narrow peak at somewhat 
higher photon energy (corresponding to $W\approx 1.726$~GeV) in $\Sigma$ for $\gamma p\rightarrow p\gamma$.
Meanwhile, a counterpart of this second peak was also established \cite{Werthmueller_15} for the 
$\gamma n\rightarrow n\eta$ reaction.

A better understanding of these experimental findings requires data beyond total cross sections and
angular distributions that can pin down the partial wave(s) related to these structures. This requires
the measurement of single- and double-polarization observables \cite{Barker_75}.   
A polarization observable that is of particular interest in the discussion of the narrow structure 
in $\eta$ photoproduction is the double-polarization observable $E$. It allows to split the results
for the unpolarized cross section $\sigma_0$ into their helicity-1/2 and 3/2 parts;
i.e., into reactions with incident photon and nucleon spins which are parallel ($\sigma_{3/2}$) or antiparallel
($\sigma_{1/2}$). This observable is defined as:
\begin{equation}
E\equiv \frac{\sigma_{1/2}-\sigma_{3/2}}{\sigma_{1/2}+\sigma_{3/2}} = \frac{\sigma_{1/2}-\sigma_{3/2}}{2\sigma_0}\, , 
\label{eq:E}
\end{equation}
and can be measured with a circularly polarized photon beam impinging on a longitudinally polarized
nucleon target. This equation can be used to extract the total asymmetry $E(W)$ when used with
total cross sections $\sigma_{1/2}(W)$, $\sigma_{3/2}(W)$ or its angular distribution $E(W,\theta^{\star})$ 
when used with differential cross sections $d\sigma_{1/2}(W,\theta^{\star})$, $d\sigma_{3/2}(W,\theta^{\star})$.
Nucleon resonances with spin $J=1/2$ appear only in $\sigma_{1/2}$, while resonances 
with spin  $J\geq 3/2$ contribute also (mostly even dominantly) to $\sigma_{3/2}$. The helicity-dependent 
cross sections therefore give insight into the spin structure of the production process.

In the present paper we present results obtained with the Crystal Ball/TAPS experiment at the Mainz MAMI
accelerator using a circularly polarized photon beam (bremsstrahlung from longitudinally polarized electrons)
and a longitudinally polarized solid deuterated butanol target. Some results for the helicity-dependent
cross sections for the quasifree $\gamma n\rightarrow n\eta$ reaction have already been published \cite{Witthauer_16}. 
Here we give a detailed account of the analysis procedures and all results for $\gamma n\rightarrow n\eta$ and 
the simultaneously investigated $\gamma p\rightarrow p\eta$ reaction with protons bound 
in the deuteron.     

\section{Experimental Setup}
\label{sec:1}

The double-polarization data were measured during four beam-time periods at the Mainz MAMI 
\cite{Herminghaus_83,Walcher_90} electron acceleration facility. The longitudinally polarized electron 
beam with an energy of $E_0\simeq1.6$ GeV was used to produce circularly polarized photons via 
bremsstrahlung tagging off an amorphous radiator (10 $\mu$m Vacoflux50). The scattered electrons were 
deflected in the magnetic field (1.9~T) of the Glasgow tagger \cite{Anthony_91,Hall_96,McGeorge_08} and 
registered in the focal plane detector composed of overlapping plastic scintillators (9 - 32~mm widths)
forming 352 logic channels of twofold coincidences. Electron energies, and the corresponding energies of the 
bremsstrahlung photons, were determined with a resolution of 2-5 MeV, which corresponds to the widths of 
the focal-plane counters. The resolution of the dipole spectrometer is much better. The tagger covers 
5 - 93\% of the incident electron energies. However, because the high count rates at low photon energies,
which were not interesting for the present experiment, would have limited the maximum usable beam current,
those sections of the focal plane detector were deactivated so that only the photon energy range  
$E_{\gamma}\approx400$ - 1450~MeV was tagged. 

The electron polarization was between $P_e\simeq80\%$ and $P_e\simeq85\%$ and was determined with the 
help of Mott scattering close to the electron source at a beam energy of 3.65 MeV \cite{Tioukine_11}. 
In addition, M{\o}ller scattering was used to monitor the electron polarization at the site of the radiator. 
The energy-dependent circular photon polarization degree, $P_{\gamma}$, follows from the polarization 
transfer formula given by Olsen and Maximon \cite{Olsen_59}:
\begin{equation}
\frac{P_{\gamma}}{P_{e}} = \frac{3+(1-x)}{3 + 3(1-x)^2 -2(1-x)}\cdot x\, ,
\label{eq:olsen}
\end{equation}
where $x=E_{\gamma}/E_0$, and $E_{\gamma}$ is the energy of the photon. The polarization degree is highest
for maximum photon energies and drops with decreasing energy. This results for the interesting energy range 
of the narrow structure in the $\gamma n\rightarrow n\eta$ excitation function ($E_{\gamma}\approx$1~GeV)
in a photon polarization degree of $P_{\gamma}\approx 0.8\times P_{e}\approx$0.66. 

The photon beam was collimated behind the radiator to a diameter of 2~mm resulting in a beam-spot size
of 9~mm on the production target, which was a longitudinally polarized, frozen-spin target \cite{Rohlof_04}. 
The target container was 2 cm long and made of Teflon. It was filled with deuterated butanol (C$_4$D$_9$OD) 
beads of 1.88~mm diameter. Dynamic nuclear polarization \cite{Bradtke_99} was used to polarize the deuterated 
butanol. The polarizing process required a magnetic field of 1.5 T and a temperature of 25 mK. 
The low temperature ensured a long relaxation time of more than 2000 hours. During data taking, the large 
polarizing magnet was exchanged for a small solenoidal holding coil with a magnetic field of 0.6 T. 
The target polarization was measured with an NMR system before and after data taking and interpolated
exponentially in between. For the first three beam-time periods, small field inhomogeneities 
($\Delta B\leq 1.78$ mT) of the polarizing magnet caused a inhomogeneous polarization across the target 
diameter. The values measured for the polarization degree with the NMR technique did therefore not
correctly reflect the polarization in the target area hit by the photon beam. This general problem was
discovered by the present experiment because the asymmetry $E$ for $\eta$ production in the threshold range
is known to be unity. The problem was investigated using a target with NMR coils which allowed separate measurements
of the polarization degree in the center and the outer layers of the target. It was solved in a
fourth beam time with a different frozen spin target. The previous targets used trityl Finland D36 as a
dopant, which produces high polarization, but has a very narrow resonance line. During the last 
beam time the older tempo dopant was used. This results in smaller polarization, but due to 
the much broader resonance line it is not sensitive to the inhomogeneities of the magnetic field. The absolute
scale of the asymmetries was rescaled to this fourth beam time.

In addition to the measurement with the solid butanol target two further beam times, one with a liquid deuterium 
target and one with a solid carbon target, were analyzed. The liquid deuterium target was used to investigate the
signal line shapes for reactions with nucleons bound in the deuteron and the measurement with the carbon target was
used to eliminate the background from the unpolarized carbon nuclei in the butanol target. The parameters of the
three targets are summarized in Table~\ref{tab:targets}.

\begin{table}[hhh]
\begin{center}
  \caption[Summary of data sets]{
    \label{tab:targets}
     Summary of targets. Target type ($SB$: solid butanol ${\rm C_4D_9OD}$, $LD_2$: liquid deuterium,  
     $C$: solid carbon foam; target length $d$ [cm]; 
     density of target material $\rho_t$; filling factor $f$, molar mass $M_{m}$ [g/mol],
     target surface number density $N_D$ [nuclei/barn] of deuterons; target surface number density
     $N_N$ [nuclei/barn] of carbon (and oxygen) nuclei.
}
\vspace*{0.3cm}
\begin{tabular}{|c|c|c|c|c|c|c|}
\hline
Target & d[cm] & $\rho_t$[g/cm$^3$] & $f$ & $M_{m}$[g/mol] & $N_D$[b$^{-1}$] & $N_N$[b$^{-1}$] \\
\hline\hline
  $SB$    & 2.0  & 1.1   & 0.6 & 84.2 & 0.094 & 0.047 \\
  $LD_2$  & 3.02 & 0.163 & 1.0 & 2.01 & 0.147 & - \\
  $C$     & 1.98 & 0.57  & 1.0 & 12.0 & - & 0.057\\    
\hline
\end{tabular}
\end{center}
\end{table} 

Since the butanol target material consisted of small beads, the target volume was not completely filled. 
The filling factor was measured to be 0.60$\pm$0.02. The solid butanol and the liquid deuterium target were of similar 
size and similar surface number density of deuterons. The carbon target was made from a special foam 
that allowed its density to be adjusted. Table~\ref{tab:targets} lists the surface number density of
nuclei in the carbon target and the surface density of carbon plus oxygen nuclei in the solid butanol target.
The density of the carbon was chosen a little higher than of the butanol because the butanol target had 
an additional 40\% filling with helium coolant and one of the nuclei in butanol is oxygen instead of carbon.
Taking into account that the photoproduction of $\eta$ mesons from nuclei scales with the nuclear mass
number $A$ like $A^{2/3}$ \cite{Roebig_96,Mertens_08} the effective surface number densities for the butanol
and carbon targets were identical. This ensured a subtraction of the nuclear background with small systematic 
uncertainties.  

\begin{figure}[t]
\centerline{
\resizebox{0.5\textwidth}{!}{\includegraphics{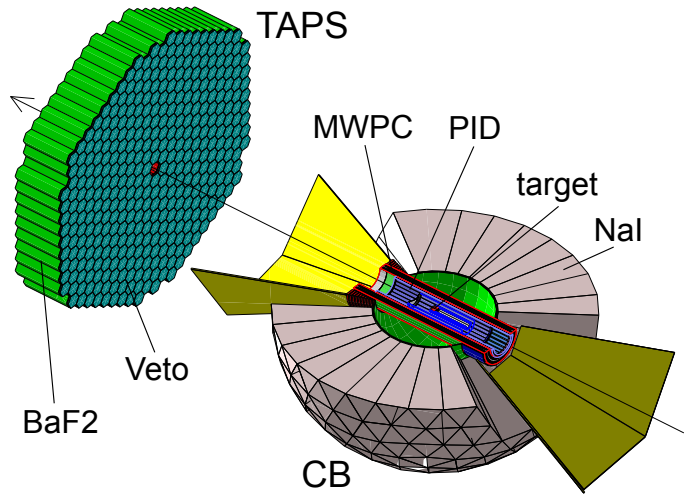}}}
\caption{Detector setup of the A2 experiment at MAMI.}
\label{fig:Setup}       
\end{figure}

The detector setup is shown in Fig.\ \ref{fig:Setup} and is described in detail in 
Refs. \cite{Witthauer_13,Werthmueller_14,Oberle_13,Oberle_14,Dieterle_15}. The main detector was an almost
$4\pi$ solid angle covering calorimeter combining the Crystal Ball detector (CB) \cite{Starostin_01} with 
the TAPS detector \cite{Novotny_91,Gabler_94}. The CB is made of 672 NaI(Tl) crystals and covered an 
angular range of $20^{\circ}\leq\theta\leq160^{\circ}$ with a typical resolution of $\Delta\theta=2-3^{\circ}$ 
and $\Delta\phi=2-4^{\circ}$. The energy resolution of the CB detector is $\Delta 
E/ E = 2\% / (E{\rm [GeV]})^{0.36}$ \cite{Starostin_01}. In the CB, charged particles were identified by
the Particle Identification Detector (PID) \cite{Watts_05}, which is made of 24 plastic 
scintillator bars with a thickness of 4 mm. A multiwire proportional chamber (MWPC) was also mounted, 
but not used for this experiment.
The TAPS detector covered the forward angular range between $\theta=5^{\circ}$ and $\theta=21^{\circ}$ 
with a resolution of $\Delta\theta\leq 1^{\circ}$ and $\Delta\phi =1^{\circ}-6^{\circ}$.
It consisted of 366 hexagonally shaped BaF$_{2}$ crystals and 72 PbWO$_4$ crystals. The photon energy 
resolution was measured to be $\Delta E / E = 1.8\% + 0.8\% / (E{\rm [GeV]})^{0.5}$ \cite{Gabler_94}. 
Each module was equipped with a 5~mm thick plastic scintillator (CPV) in front of the BaF$_{2}$ crystal, 
which was used for charged particle identification. 

The experimental trigger required at least two activated detector clusters in the combined system.
For this purpose, TAPS was divided into six triangular logic sectors. A TAPS sector contributed to the total 
multiplicity if at least $\sim 35$ MeV were deposited in one detector module of the sector. 
Analogously, the CB detector was divided into sectors of 16 adjacent crystals each, the energy in one 
sector had to be above 10 - 30~MeV to add to the total multiplicity. In addition, events from single pion 
production from the $\Delta$-region were suppressed by requiring an energy deposition (analog sum of the 
energy signals) of at least 250~MeV in the CB detector.

\section{Data Analysis}
\label{sec:Ana}
The primary data analysis, i.e. the identification of $\eta$ mesons from their
$\eta\rightarrow\gamma\gamma$ and $\eta\rightarrow 3\pi^0\rightarrow 6\gamma$ decays and the identification 
of recoil nucleons was analogous to the one described in Refs.~\cite{Witthauer_13,Werthmueller_14} and will 
only be briefly summarized. Also the identification for reactions off nucleons bound in 
the deuteron, e.g. suppression of background from multiple pion production, with coplanarity and missing-mass
analyses was identical to the methods described in Refs.~\cite{Witthauer_13,Werthmueller_14}. The additional 
background from reactions with nucleons bound in the carbon (and oxygen) nuclei produces broader 
structures in these spectra and cannot be completely suppressed. This background cancels for the numerator in 
Eq.~\ref{eq:E} because these nucleons are not polarized. For the denominator, one can either use the results
from measurements with a liquid deuterium target or one must subtract the nuclear background measured with 
a solid carbon target.
   
\subsection{Particle and Reaction Identification}
\label{sec:ID}
In the first step of the analysis, clusters of activated crystals were searched in the CB and in TAPS
and assigned with the help of the PID and CPV scintillators to the two lists of `neutral' and `charged' 
hits in the calorimeters. Based on the number of `charged' and `neutral' clusters, events were attributed 
to one of the four classes listed in Table \ref{tab:EvC}. Events outside these classes were rejected 
from the analysis to reduce background contributions. 
\begin{table}[h]
\begin{tabular}{|c|c|c|}
\hline
$\eta$ decay mode & reaction & criteria \\ \hline
$\eta \to 2 \gamma$ & $\gamma p \to p \eta $ & 2n and 1c \\
$\eta \to 2 \gamma$ & $\gamma n \to n \eta $ & 3n \\
$\eta \to 6 \gamma$ & $\gamma p \to p \eta $ & 6n and 1c \\
$\eta \to 6 \gamma$ & $\gamma n \to n \eta $ & 7n \\ \hline
\end{tabular}
\caption{Analyzed event classes. `n' incidcates `neutral' hits, `c' `charged' hits.}
\label{tab:EvC}  
\end{table}

The photons from the $\eta\to2\gamma$ and the $\eta\to3\pi^0\to6\gamma$ decay were registered in 
coincidence with the recoil nucleon, i.e. in an exclusive measurement. For events with one charged cluster,
this cluster was directly assigned to the recoil proton. For events with only two neutral hits, the invariant
mass of those two hits (assuming that they were photons) was compared to the invariant mass of the $\eta$ meson.
For events with more than two neutral clusters, a $\chi^2$ test was performed for the invariant masses
$m_{\gamma\gamma}$ of all combinatorial possible partitions of the neutral hits to pairs. For events with 
three neutral hits, the invariant masses were compared to the nominal mass of the $\eta$ meson 
($m_{\eta}$=547.862~MeV \cite{PDG_16}) using 
\begin{equation}
\chi^2 =  \left( \frac{m_{\gamma\gamma}-m_{\eta (\pi^0)}}{\Delta m_{\gamma\gamma}}\right)^2
\label{eq:Chi1}
\end{equation}
and also to the mass of the $\pi^0$ meson ($m_{\pi^0}$=134.9766~MeV \cite{PDG_16}). 
In Eq.~\ref{eq:Chi1}, $\Delta m_{\gamma\gamma}$ represents
the uncertainty due to experimental resolution of the measured invariant masses, which was determined with 
Monte Carlo (MC) simulations. Events from this class for which the smallest $\chi^2$ corresponded to the $\pi^0$ 
invariant mass were discarded to reduce background. For the other events with three neutral clusters, the 
hits from the `best' combination of neutral pairs to the $\eta$ invariant mass were assigned as photons and 
the remaining bachelor hit was assigned as the recoil neutron. In a similar way, hits from events with 
six or seven neutral clusters were tested against the invariant mass of the $\pi^0$ meson using  
\begin{equation}
\chi^2 =  \sum_{i=1}^{3}\left( \frac{m_{\gamma\gamma}-m_{\pi^0}}{\Delta m_{\gamma\gamma}}\right)^2\, .
\label{eq:Chi2}
\end{equation}

For events with seven neutral clusters, again the hit not assigned as a meson decay photon 
was identified as recoil neutron. Furthermore, for events with six or seven neutral hits a $\chi^2$ test 
was also used to assign the photons pairwise to their parent pions.
This assignment helps to improve the resolution for the following analysis steps because the energies
for each pair of photons from a $\pi^0$ decay can be recalibrated using the nominal mass of the
$\pi^0$ by  
\begin{equation}
E'_{\gamma_1,\gamma_2} = \frac{m_{\pi^0}}{m_{\gamma_1\gamma_2}} \cdot E_{\gamma_1,\gamma_2}\, ,
\label{eq:xform}
\end{equation}
where $E_{\gamma_1,\gamma_2}$ are the measured energies and $E'_{\gamma_1,\gamma_2}$ are the recalibrated 
ones. This correction is based on the fact that the angular resolution of the calorimeter is much better 
than the energy resolution, so that most of the deviation between the measured invariant mass
$m_{\gamma_1,\gamma_2}$ and the nominal pion mass $m_{\pi^0}$ is due to the photon energy measurement.
The same correction was applied to the two-photon events using the $\eta$ mass for recalibration.

The combinatorial $\chi^2$ analysis described above can be applied to all hits in the calorimeter no 
matter whether they were detected in the CB-PID or TAPS-CPV system. Further particle identification 
methods were available individually for the two detector systems and were used to cross check the
correct assignment of all hits, as discussed in detail in \cite{Werthmueller_14}. 

In TAPS, a clean separation of neutrons from photons was possible with a pulse-shape analysis (PSA) of the 
two scintillation light components of the BaF$_2$ crystals, as described in \cite{Werthmueller_14}. 
Furthermore, time-of-flight versus energy was also used to separate photons from massive particles
in TAPS. 

In the CB, $E-\Delta E$ spectra using the CB-PID system allow a clean separation of protons from charged 
pions. In this system, an analysis of the cluster multiplicity (i.e. the number of modules activated per
cluster) can be used to cross check the correct separation of neutrons from photons, because the 
electromagnetic showers from photon hits spread over a larger number of modules than hits from neutrons.
Altogether, as shown in Ref.~\cite{Werthmueller_14}, the combination of these methods allows a very clean
identification of photons, protons, and neutrons in the CB/TAPS detector system. 

\begin{figure*}[h]
\centerline{
\resizebox{0.8\textwidth}{!}{\includegraphics{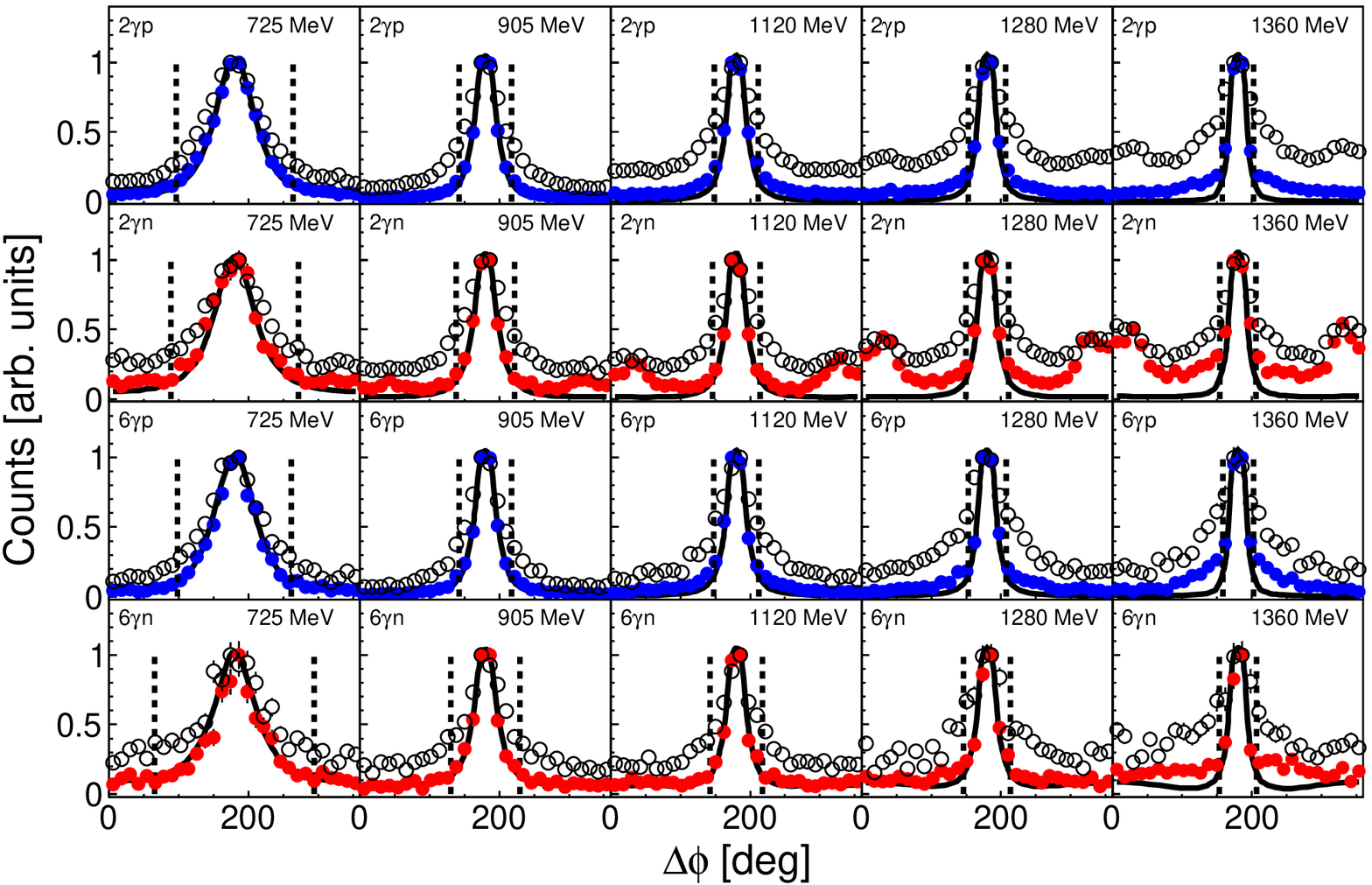}}}
\caption{(Color online) Coplanarity spectra. The angular difference $\Delta\phi$ between the recoil nucleon and 
the $\eta$ meson for five different bins of incident photon energy. The spectra are integrated over 
the whole angular range and were filled right after the $\chi^{2}$ selection, the PSA and the 
invariant-mass cut were applied. The results for the deuterium target are shown in colors (red and 
blue solid circles) and the results for the deuterated butanol target are shown as open black circles. 
The MC line shape is shown as a solid black line. The dashed lines show the $2\sigma$ 
cut position determined from the simulation.}
\label{fig:Cop}       
\vspace{0.5cm}
\centerline{
\resizebox{0.8\textwidth}{!}{\includegraphics{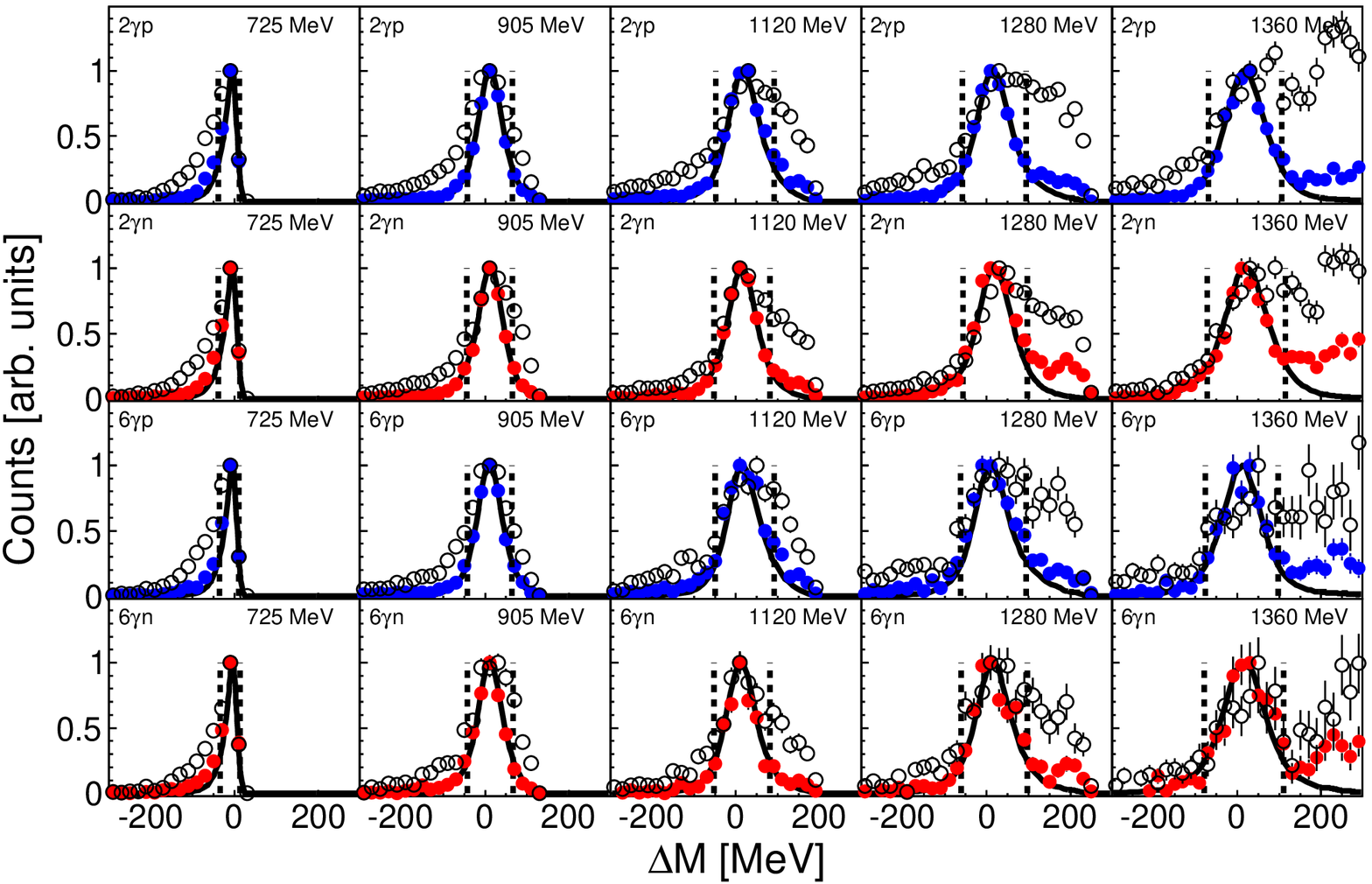}}}
\caption{(Color online) Missing mass $\Delta M$ for five different bins of incident photon energy. 
The spectra are integrated over the whole angular range and were filled after the $\chi^{2}$ selection, the PSA, 
the coplanarity and the invariant-mass cut were applied. Shown are the results for the $\eta\to2\gamma$ 
(first two rows) and $\eta\to6\gamma$  decay (last two rows). The results for the deuterium target 
are shown in colors (red and blue solid circles) and the results for the deuterated butanol target are 
shown as open black circles. The cut position of $\pm 1.5 \sigma$ is indicated by the dashed line.}
\label{fig:MM}       
\end{figure*}
\begin{figure*}[tb]
\centerline{
\resizebox{0.85\textwidth}{!}{\includegraphics{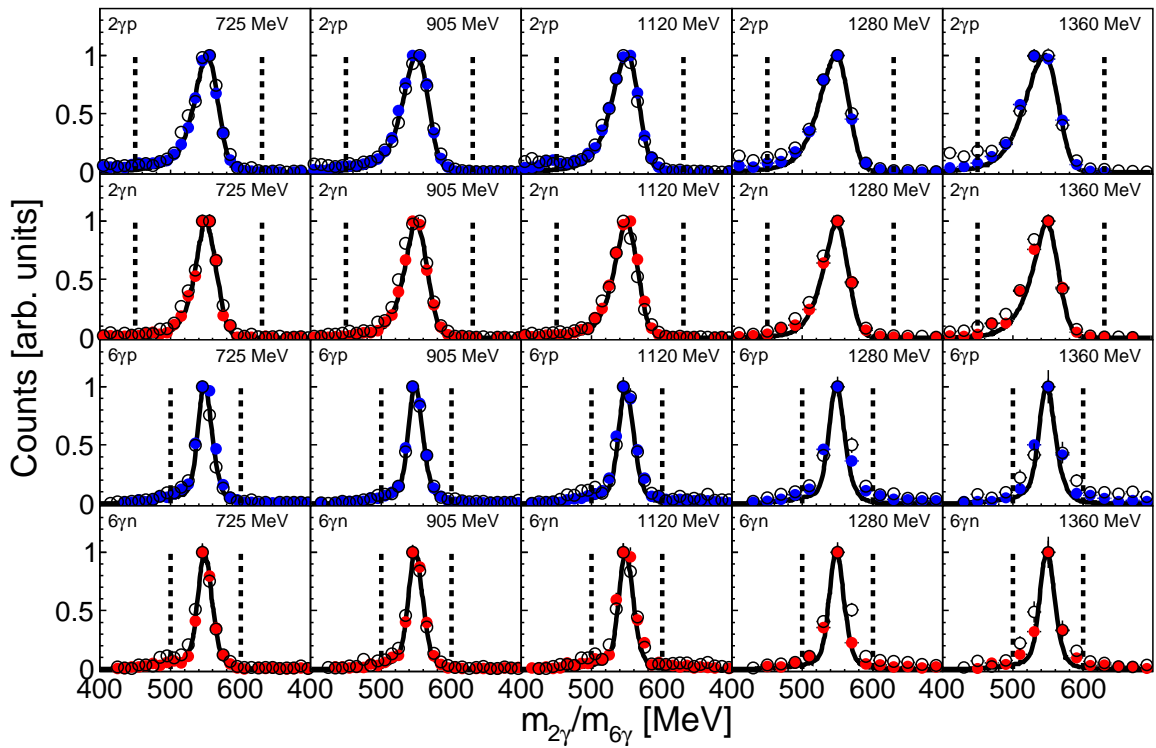}}}
\caption{(Color online) Invariant mass for five bins of incident photon energy. The spectra are integrated 
over the whole angular range and were filled after all analysis cuts (PSA, coplanarity, missing mass) were 
applied. Shown are the results for the $\eta\to2\gamma$ (first two rows) and $\eta\to6\gamma$  decay 
(last two rows). The results for the deuterium target are shown in colors (red and blue solid circles) and 
the results for the deuterated butanol target are shown as open black circles. The result of the MC
simulation is shown as solid black line. The cut positions are indicated as dashed lines.}
\label{fig:IM}       
\end{figure*}

After the hit identification and event selection, background from competing reactions was suppressed
with several analyses of the reaction kinematics. The first analysis was for the coplanarity of the $\eta$
meson and the recoil nucleon. The azimuthal angle of the $\eta$ was reconstructed from the three vectors
of its decay photons and compared to the azimuthal angle of the recoil nucleon. In the center-of-momentum
(c.m.) frame, and also in the laboratory frame, the difference between the two azimuthal angles must be 180$^{\circ}$.
Corresponding spectra for a liquid deuterium target and the butanol target are shown in Fig.~\ref{fig:Cop}
for recoil protons and recoil neutrons and for the $2\gamma$ and $6\gamma$ decay of the $\eta$ mesons.
The line shape for the measurement with the deuterium target was reproduced with a MC simulation
of the reaction taking into account the momentum distribution of nucleons bound in deuterium. 
For the events with three neutral hits (assumed $2\gamma n$), in particular at higher incident photon energies,
background is visible that peaks at azimuthal angular differences close to zero and 360$^{\circ}$. 
This background is mainly due to $\pi^0 n\to2\gamma n$ reactions where one photon was mixed up with the neutron, which
accidentally generated an invariant mass close to the $\eta$ mass. For events with recoil protons, background
comes mainly from reactions with charged pions, e.g. from the $\eta\pi^+$ final state when the $\pi^+$
was misidentified as a proton. Such backgrounds were removed with the subsequent missing-mass analysis. 

The line shape for the butanol target was broader due to the background from reactions on nucleons from 
the heavier target nuclei, which have larger Fermi momenta and are affected by FSI.
Cuts were applied at $2\sigma$ around the peak position for the deuterium target. Identical cuts were
applied to the data from the butanol target. The cuts indicated in Fig.~\ref{fig:Cop} are only schematic, 
because these spectra are integrated over angles. In the analysis, the $2\sigma$ cuts were applied 
individually for each bin of incident photon energy and of $\cos({\theta_{\eta}^{\ast}})$, where 
$\theta_{\eta}^{\ast}$ is the $\eta$ c.m.\ polar angle.    

Subsequently, a missing-mass analysis was used to remove residual background in particular
from photoproduction of $\eta\pi$ pairs, which can evade all previous selection steps when, for example, 
low-energy charged pions escape detection. If the initial-state nucleon is assumed to be at rest (Fermi motion will only
broaden the peak structure), the mass of the recoil nucleon can be deduced from the kinematics of the $\eta$:
\begin{equation}
M = \sqrt{\left(E_{\gamma}+ m_N -E_{\eta}\right)^2-\left(\vec{p}_\gamma - \vec{p}_\eta \right)^2}\, ,
\label{eq:MM}
\end{equation}
where $E_{\gamma}$ and $\vec{p}_{\gamma}$ are the energy and momentum of the incident photon beam, 
$E_{\eta}$ and $\vec{p_{\eta}}$ are the energy and momentum of the $\eta$ meson, and $m_N$ is the 
nucleon mass. Subtracting the nucleon mass from Eq.\ \ref{eq:MM} yields the missing mass $\Delta M$, 
which must peak around zero for single $\eta$ production. Typical spectra are summarized in 
Fig.~\ref{fig:MM}, the actual analysis was again done in bins of incident photon energy and $\eta$ c.m.\
polar angle. Shown are the results for the deuterium target (colored symbols), the deuterated butanol 
target (open black circles), and the MC simulation for the deuterium target (black solid line). 
The Fermi motion causes an asymmetric shape of the peak close to threshold, since Fermi momenta in 
the negative $z$-direction lead to higher c.m. energies and are thus favored. Fermi momentum and 
FSI effects are clearly more apparent in the deuterated butanol spectra than in the deuterium spectra 
due to the carbon contribution. With increasing energy, the contamination from the $\eta\pi$ reaction 
accumulates at positive missing-mass values. This background was sufficiently rejected with a cut at 
$1.5\sigma$. As for the coplanarity cut, the cut positions (dashed lines) were determined for bins of 
incident photon energy and $\cos{(\theta_{\eta}^{\ast})}$ from the deuterium data.

The reaction yields, finally used for the extraction of cross sections, were determined from the analysis
of the $\eta$ invariant-mass spectra after the application of all other cuts, in particular coplanarity
and missing mass. Typical invariant-mass spectra for the $2\gamma$ and $6\gamma$ decays of $\eta$ mesons
measured in coincidence with recoil protons and recoil neutrons are shown in Fig.~\ref{fig:IM}.
The line shapes were almost identical for the liquid deuterium and butanol target and agreed well with
the results of MC simulations. The peaks were more narrow for the $\eta\rightarrow 6\gamma$ decay
than for $\eta\rightarrow 2\gamma$ because for the latter the recalibration of photon energies using the
nominal mass of the intermediate pions with Eq.~\ref{eq:xform} improved the energy resolution. The line
shapes did not vary significantly with incident photon energy or $\eta$ c.m.\ polar angle. The integration 
of the yields was therefore done for all bins of $E_{\gamma}$ and cos$(\theta_{\eta}^{\star})$ for
the same range of $\eta$ invariant masses. This range was chosen as 450 - 620~MeV for the
$\eta\rightarrow 2\gamma$ decay and between 500 - 600~MeV for the $\eta\rightarrow 6\gamma$ decay.
There is no significant background visible in the invariant-mass spectra, but for the butanol target
these spectra include background from quasifree $\eta$ production off carbon (oxygen)
nuclei, which is indistinguishable in invariant mass and not completely suppressed by the previous
missing-mass analysis (see Sec.~\ref{sec:ExO}). 

\subsection{Extraction of the Observables}
\label{sec:ExO}

The aim of the measurement was the extraction of the polarization observable $E$ and the helicity-dependent 
cross sections $\sigma_{1/2}$ and $\sigma_{3/2}$ for parallel and antiparallel orientation
of photon and nucleon spin. Ideal results would be for free protons and free neutrons. Practically, for 
neutrons one can only measure with the quasifree nucleons bound in the deuteron. However, at least the
effects from nuclear Fermi motion can be almost completely removed by a full kinematic reconstruction
of the final state, which allows to recover the `true' c.m.\ energy $W=\sqrt{s}$ of the $\eta$ - nucleon
system. This method was discussed in detail in \cite{Werthmueller_14}. It uses the four momenta of the
meson $p_{\eta}$ and the recoil nucleon $p_{N}$ to construct $W$ via
\begin{equation}
W = \sqrt{p_{\eta}^2 + p_{N}^2}\;.
\end{equation} 
The four momentum of the $\eta$ follows directly from the measured momenta of its decay photons.
For recoil neutrons only the polar and azimuthal angles, i.e., the direction of their momenta, are measured.
The kinetic energy is unknown. Together with the three momentum of the spectator nucleon,
four kinetic observables are unmeasured. Since the incident photon energy
and the masses of all involved particles are also known, the four missing variables can be reconstructed from 
the four constraints following from energy and momentum conservation. Therefore, results can be given both
as a function of the measured incident photon energy (these are folded with Fermi motion) and as a function
of the reconstructed $W$, which are not influenced by Fermi motion. 

\begin{figure}[t]
\centerline{
\resizebox{0.5\columnwidth}{!}{\includegraphics{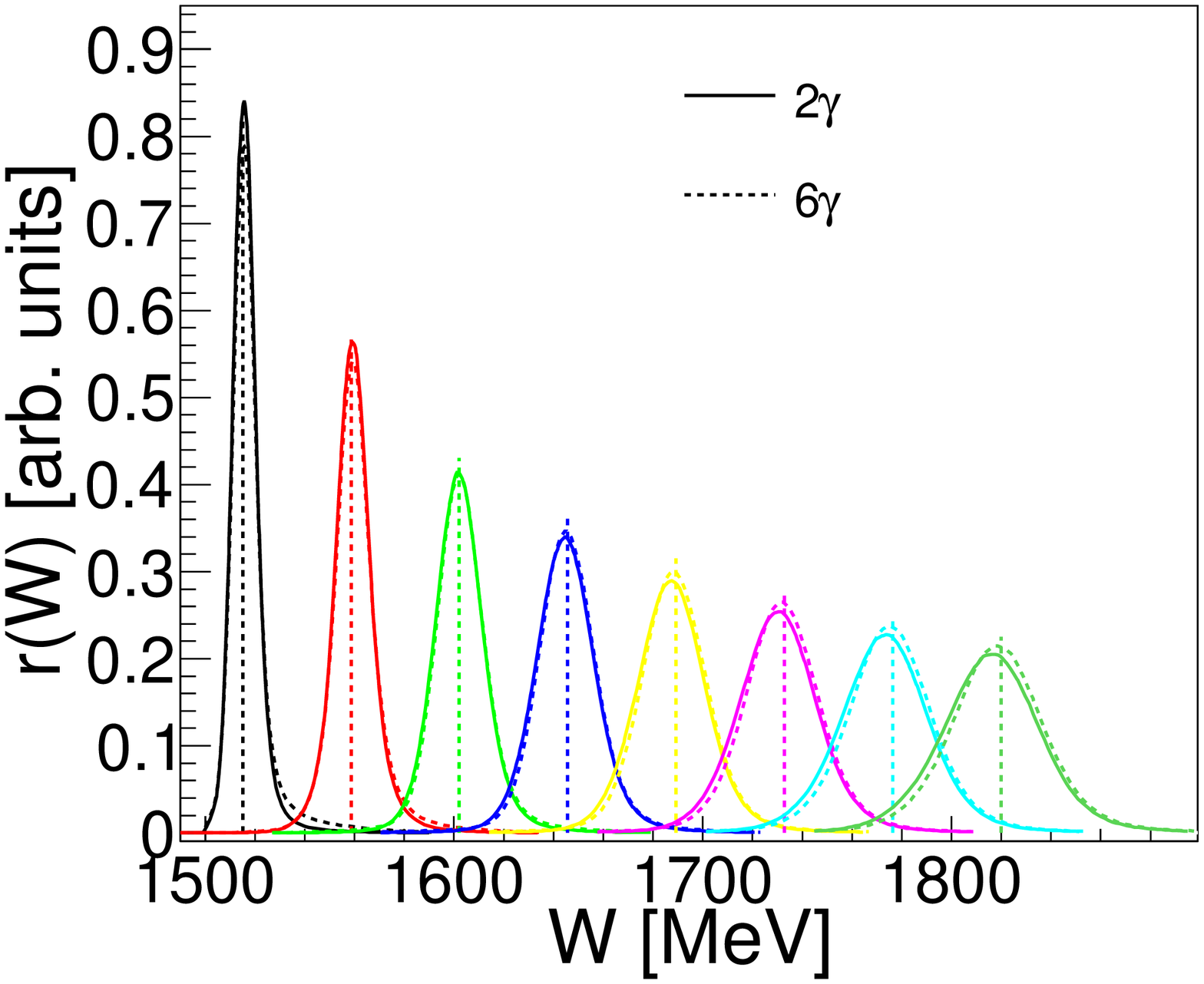}}
\resizebox{0.5\columnwidth}{!}{\includegraphics{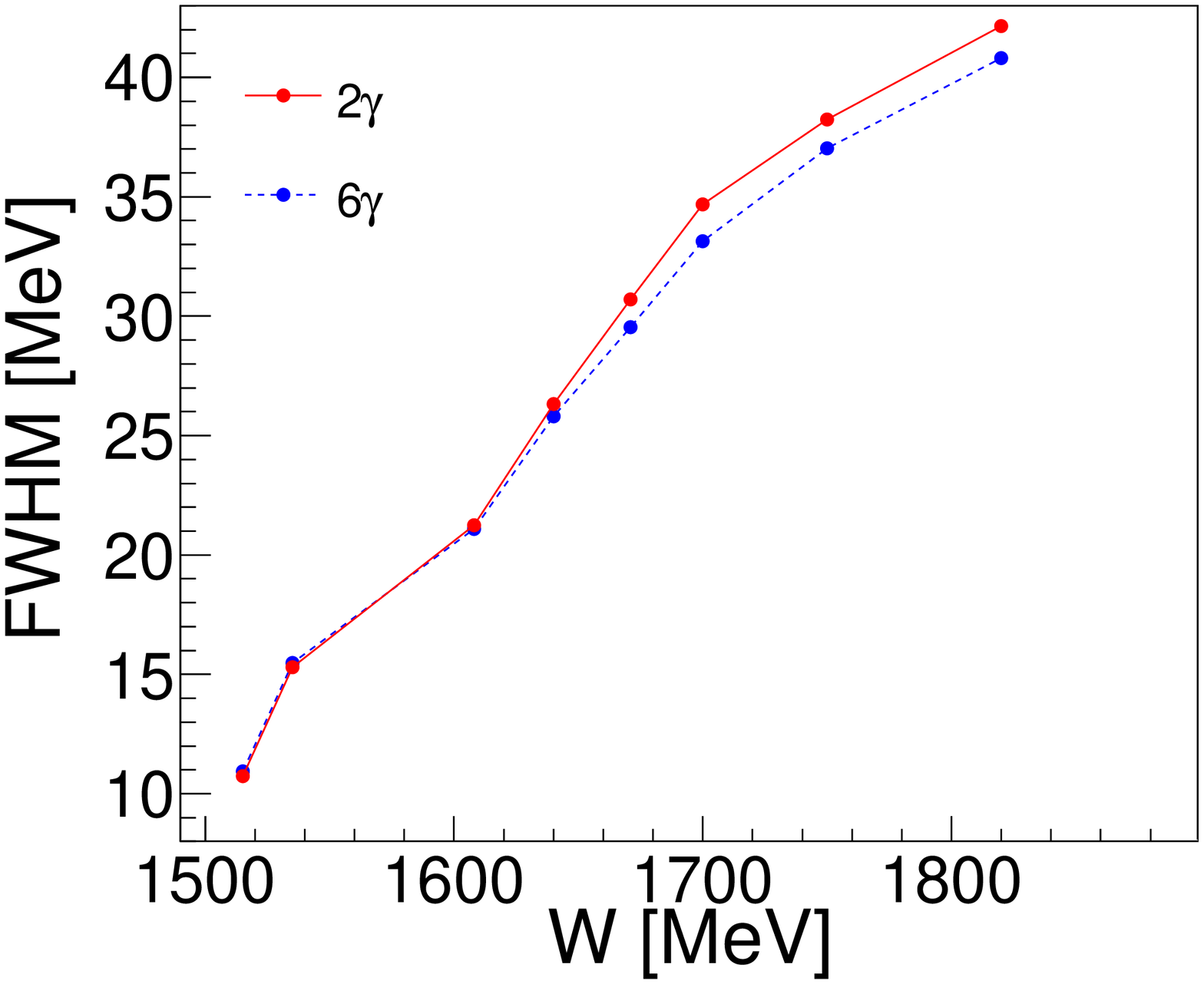}}
}
\caption{(Color online) Left-hand side: simulated response of the detection system to fixed values of $W$ (vertical dashed lines)
for the $\eta\rightarrow 2\gamma$ and $\eta\rightarrow 6\gamma$ decays. Right-hand side: FWHM of the response
as function of $W$.}
\label{fig:resol}       
\end{figure}

The $W$ reconstructed results are, however, subject to effects from experimental resolution because the 
measured $\eta$ three momenta and the polar and azimuthal angles of the recoil nucleons are used in the 
reconstruction. The resolution has been determined with a full MC simulation of the detector response
\cite{Werthmueller_14}. Phase-space distributed events were generated for several fixed values of $W$, the
events were tracked through the detector with the Geant4 code \cite{GEANT4} and analyzed like the experimental
data. The results are shown in Fig.~\ref{fig:resol}. Both $\eta$-decay modes have nearly identical resolutions
with that for $\eta\rightarrow 6\gamma$ decays slightly better than for $\eta\rightarrow 2\gamma$ decays at
higher energies. This is a bit counter intuitive, but can be easily understood, using the constraints from
the invariant mass of the mesons. The three constraints from the $\pi^0$ invariant masses for the 
$\eta\rightarrow 3\pi^0$ decay correct the energies slightly better than the one constraint from the $\eta$ 
mass for the $\eta\rightarrow 2\gamma$ decay. In the main region of interest, around the narrow structure, 
the resolution is $\approx 30$~MeV. This means that the natural width of the structure is even more narrow 
than it appears, for example, in Fig.~\ref{fig:Tot}. This has been quantitatively investigated in Ref. 
\cite{Werthmueller_14}.

It was demonstrated in \cite{Werthmueller_14} by a comparison of results measured for free protons 
(hydrogen targets) and quasifree protons bound in the deuteron that in the energy range of interest FSI effects 
are negligible for $\eta$ photoproduction. This means that the $W$ reconstructed results represent a close 
approximation of the free $\gamma n\rightarrow n\eta$ reaction. For the quasifree $\gamma p\rightarrow p\eta$ 
reaction, the kinetic energy of the recoil proton is available from the response of the calorimeter. However, 
in order to reduce systematic effects in the comparison of reactions with recoil protons and recoil neutrons,
it was not used for the $W$ reconstruction, but the reconstruction was done analogous to the neutron case 
using only the angles.   

\begin{figure}[t!]
\centerline{
\resizebox{0.9\columnwidth}{!}{\includegraphics{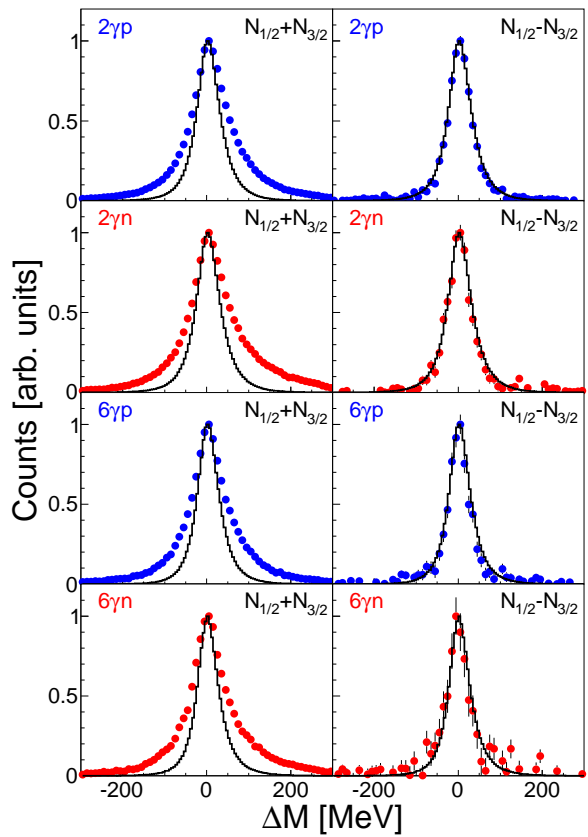}}}
\caption{(Color online) Missing mass $\Delta M$ for deuterated butanol for the difference, $N_{1/2}-N_{3/2}$, 
and the sum, $N_{1/2}+N_{3/2}$, of the two helicity states for the reaction on the proton (blue) and the neutron 
(red). The line shape of the simulation is shown as a black line. The influence of the carbon is clearly 
visible in the sum, whereas for the difference, the simulation and the experimental data are in 
agreement. The spectra are integrated over all incident photon energies and are thus dominated by the 
count rates from the $N(1535) 1/2^-$ region.}
\label{fig:MMDiff}       
\end{figure}

The measurement of an asymmetry usually does not require an absolute calibration of the reaction yields.
However, due to the background from reactions with unpolarized nucleons bound in the heavier nuclei of
the butanol target this is different here. The effect is demonstrated with the missing-mass spectra
shown in Fig.~\ref{fig:MMDiff}. The left hand side of the figure shows missing-mass spectra for the sum 
of the yields for the two relative spin orientations $N_{1/2}$ and $N_{3/2}$ after all other cuts,
the right hand side the difference of the same quantities. The experimental results are compared to the 
MC-simulated line shape for quasifree production from a deuteron target. The agreement is good for the 
difference of the count rates, for which all unpolarized contributions cancel, but the sum  
includes unpolarized nuclear background that involves larger Fermi momenta. Note that the background 
due to other reaction channels, in particular production of $\pi\eta$ pairs, visible in Fig.~\ref{fig:MM} 
appears much less prominent in Fig.~\ref{fig:MMDiff} because the spectra are integrated over photon energy 
and thus dominated by the $N(1535) 1/2^-$ signal, which is not contaminated with $\eta\pi$ background.  

There are two different methods to eliminate this background from the denominator of Eq.~\ref{eq:E}.
Both methods use in the numerator of Eq.~\ref{eq:E} the difference of the $\sigma_{1/2}$ and $\sigma_{3/2}$
cross sections measured with the butanol target. One method, which we call version (1), uses in
the denominator for $\sigma_{1/2}+\sigma_{3/2}$ the results from the butanol target after subtraction of
the unpolarized background measured with the carbon foam target. In the other method, version (2), the 
denominator is replaced by $2\sigma_{0}$, where $\sigma_{0}$ is the unpolarized cross section measured with
a liquid deuterium target. Both methods require, however, that the asymmetry is not simply constructed
from uncalibrated count rates but from absolutely calibrated cross sections because both combine two 
measurements with different targets, different photon fluxes, and some other different experimental settings.
For this experiment version (1) has smaller systematic uncertainties because the experimental conditions for 
the measurements with the butanol and the carbon target were very similar. They had the same target size, 
same target density, same target containment, same experimental conditions in view of trigger conditions, etc. 
and were measured shortly one after the other. The measurement with the liquid deuterium target was done much 
earlier, the target had a different size and density, the target environment was different and also some other 
experimental details had been modified between these measurements. Therefore, for the comparison of 
butanol and carbon target data, many experimental factors cancel so that mainly the well measured photon fluxes 
had to be eliminated. Other factors like detection efficiencies, target thickness etc. were also 
taken into account but played a minor role. On the other hand, for the combination of butanol and liquid 
deuterium data in Eq.~\ref{eq:E}, exact absolute normalizations taking into account all experimental variables 
were mandatory.

\begin{figure}[t]
\centerline{
\resizebox{\columnwidth}{!}{\includegraphics{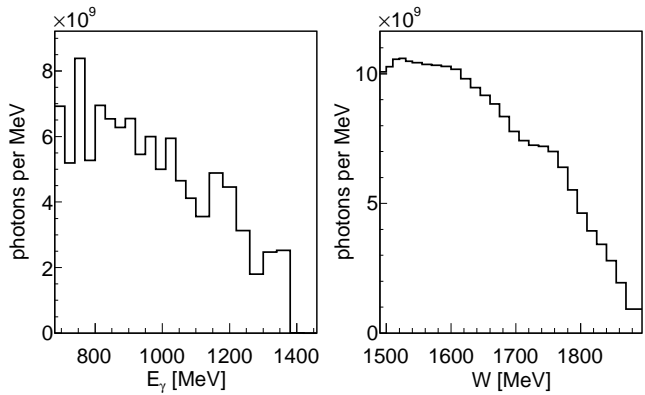}}}
\caption{Incident photon flux, i.e. count rate of scattered electrons times tagging efficiency for the 
measurement with the butanol target. Left-hand side: as function of photon energy measured with the tagging
spectrometer. Right-hand side: as function of reconstructed $W$ after folding with Fermi motion.}
\label{fig:Flux}       
\end{figure}

For the measurements with all three targets, absolute cross sections were derived from the extracted yields,
the decay branching ratios for the $\eta\rightarrow 2\gamma$ (39.41$\pm$0.20)\% and the $\eta\rightarrow 6\gamma$
(32.68$\pm$0.23)\% \cite{PDG_16} decays, the target densities, the measured photon fluxes, and the 
simulated detection efficiencies.

The photon flux was derived from the number of scattered electrons, counted with the scalers of the tagger
focal-plane detectors, and the tagging efficiency, i.e.\ the fraction of bremsstrahlung photons that pass the 
collimator and irradiate the target. The tagging efficiency was measured absolutely approximately once per day 
with dedicated low intensity runs for which a $\approx100$\% efficient lead glass detector was moved into the 
photon beam. The relative stability of the tagging efficiency between those measurements was monitored in the 
offline analysis with the help of the yield from the $\gamma X\rightarrow X\pi^0$ reaction. Typical values of
the tagging efficiency for the butanol measurements were in the 30\% range. The photon flux derived from this
analysis can be directly applied to the data measured as a function of incident photon energy. For the analysis
as a function of reconstructed $W$, it must be folded with the momentum distribution of nucleons bound in the
deuteron taken from \cite{Lacombe_81}. The two flux distributions are shown in Fig.~\ref{fig:Flux}.
The difference in shape between the flux distributions as functions of $E_{\gamma}$ and $W$ and the disappearance
of the fluctuations for $W$ comes respectively from the folding with Fermi motion and the change in scale from the 
Jacobian in the transformation from $E_{\gamma}$ to $W$.

\begin{figure*}[t]
\centerline{
\resizebox{1.0\textwidth}{!}{\includegraphics{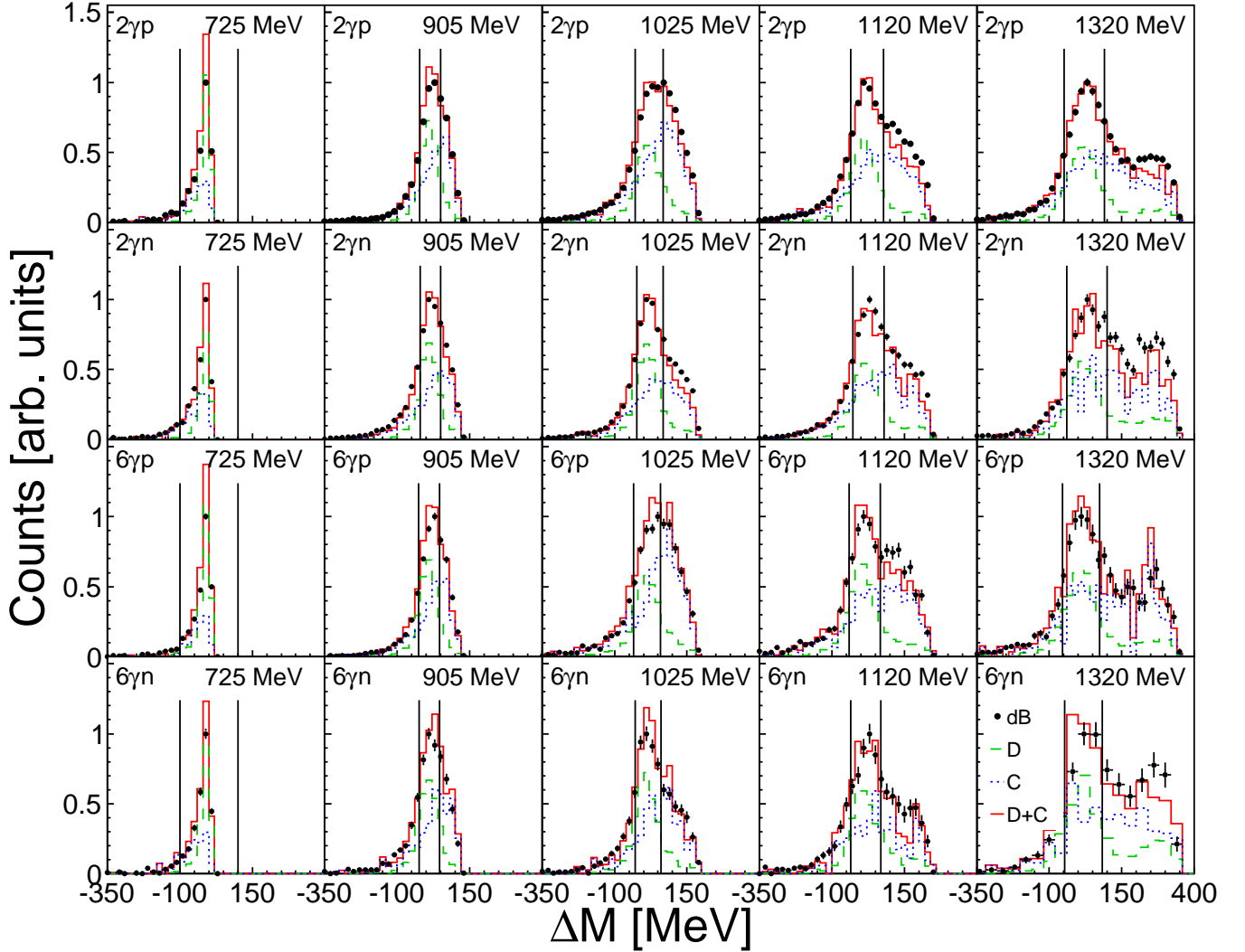}}}
\caption{(Color online) Missing-mass contribution from the deuterium target (dashed green line), the contribution from 
the carbon target (dotted blue line), and the deuterated butanol distribution for $\sigma_{\rm sum}$ (black dots). 
The sum of the deuterium and the carbon is shown in red. The yields from the different targets were absolutely 
normalized with the target densities, the fluxes, and the detection efficiencies. Only the overall scale of 
the figures is in arbitrary units. A variable energy binning was used (mean value indicated) and only a 
selection of bins is shown here.}
\label{fig:FitMM}       
\end{figure*}

The main tool for the determination of the detection efficiency was Monte Carlo simulations with the Geant4 \cite{GEANT4}
code. Detailed results for the measurement with the liquid deuterium target were shown in \cite{Werthmueller_14}.
These simulations are very well tested and reliable for the electromagnetic showers from the meson decay photons,
but less so for the recoil nucleons. In particular, low energy neutrons and protons passing the inactive support
structures in the transition region from the CB to the TAPS detector are critical. For the measurement of the
unpolarized cross sections \cite{Werthmueller_14}, such effects were studied in detail and corrected by the 
analysis of data obtained with a liquid hydrogen target. Correction factors for the detection of recoil protons 
and recoil neutrons were determined by the analysis of the $\gamma p\rightarrow p\eta$ and 
$\gamma p\rightarrow n\pi^0\pi^+$ reactions \cite{Werthmueller_14} as a function of recoil nucleon laboratory polar 
angle and kinetic energy, where the latter was reconstructed from reaction kinematics. Such corrections were also applied 
for the butanol target, but they are less precise in this case because the hydrogen target had a different 
material budget (important for low energy protons) and the hydrogen data were measured long before the butanol 
data under not identical experimental conditions. This is the main reason why the extraction of $E$ using 
$2\sigma_0$ in the denominator of Eq.~\ref{eq:E} has a larger systematic uncertainty than the carbon subtraction.     

The target densities are given in Table~\ref{tab:targets}. The comparison of contributions from deuterons in the 
butanol and liquid deuterium targets is straightforward. For the comparison of the contributions from carbon, 
oxygen, and helium nuclei in the butanol target to the yields measured with the carbon foam target, one must not
only take into account the surface number densities of the targets but also the scaling of the $\eta$ yields
with $A^{2/3}$ \cite{Roebig_96, Mertens_08}. The effective surface number densities taking into account these
effects are 0.0376 (C), 0.0114 (O), and 0.008 (He) (sum = 0.057) for the butanol target and 0.057 for the carbon
target (all in units of 1/barn). One should note that the spectral distributions for quasifree
$\eta$ production of nucleons bound in carbon and helium nuclei are similar, because the larger FSI effects in 
carbon are counteracted by larger Fermi momenta in helium.

Finally, to arrive at helicity cross sections, the data have to be normalized by the target and beam 
polarizations discussed in Sec.~\ref{sec:1}.      
  
After the normalizations have been applied to the data, one can compare the missing-mass spectra
for the three different targets obtained after all other experimental cuts. This is shown in Fig.~\ref{fig:FitMM}.
It should be emphasized that the relative normalization of the three yields has no free parameter, only the 
absolute scale in the figure is arbitrary. The data measured with the liquid deuterium and carbon foam target
nicely add up to the measurement with the butanol target. At higher incident photon energies, some deviations 
occur in the background region of the spectra. This may be due to larger differences for $\eta\pi$ pairs
than for single $\eta$ production in the spectral shapes for the production off carbon and helium nuclei. 
It does, however, not matter here because it only affects the behavior in the background region (the agreement 
in the peak region was much better) and, due to the absolute calibration of cross sections, the background 
region was not used for normalization purposes.    

At very low energies there is, in particular for the proton data, a discrepancy 
between butanol data and the sum of carbon and liquid deuterium data. This can be traced to a problem with the 
detection efficiency for recoil protons for the measurements with the butanol and carbon targets, which is 
discussed below. The liquid deuterium data are, for version (1) of the analysis, only used for the cross check 
that deuterium and carbon data add up to the butanol data. The yields for this analysis are determined by the 
difference of the butanol and the carbon data, the liquid deuterium data are not needed for this extraction.

Primarily extracted from the butanol and carbon data were two sets of differential cross sections defined by:
\begin{eqnarray}
\frac{d\sigma_{\rm diff}}{d\Omega} & = & \frac{d\sigma_{1/2}}{d\Omega} - \frac{d\sigma_{3/2}}{d\Omega}\nonumber\\
\frac{d\sigma_{\rm sum}}{d\Omega}  & = & \frac{d\sigma_{1/2}}{d\Omega} + \frac{d\sigma_{3/2}}{d\Omega}\, .
\end{eqnarray} 
The cross section with label `diff' represents the difference of the helicity-1/2 and helicity-3/2 components
from the butanol target for which unpolarized carbon background cancels automatically. The unpolarized carbon 
background was explicitly subtracted for the `sum' cross section. The total cross sections $\sigma_{\rm diff}$
and $\sigma_{\rm sum}$ have been determined with fits of Legendre polynomials to the differential ones.  

The total and differential asymmetries $E$ were then constructed in the two different ways discussed above,
i.e. either as $\sigma_{\rm diff}/\sigma_{\rm sum}$ or as $\sigma_{\rm diff}/2\sigma_0$, where the 
unpolarized cross section $\sigma_0$ was taken from the measurement with a liquid deuterium target
\cite{Werthmueller_14}. In the latter version, unpolarized background cancels in the numerator and is 
not present in the denominator. However, this analysis is more dependent on an exact absolute normalization 
of the butanol data because experimental conditions were different from the measurement with the liquid 
deuterium target. The main problem for the absolute calibration of the butanol as well as the carbon data 
is the detection efficiency for recoil protons that were detected close to the transition region between 
CB and TAPS. In this region are holding structures that the particles pass through and which are not precisely
described in the MC simulations. 
In contrast to the measurement of the unpolarized cross section \cite{Werthmueller_14}, there were no data 
available to extract precise correction factors for these effects. They were in particular important for 
the energy range from threshold throughout the $N(1535) 1/2^-$ resonance region. This imperfect 
detection efficiency correction leads to incorrect absolute cross sections for the reaction with quasifree
protons at low energies. The proton results for the $E$ asymmetry from analysis (2) are therefore discarded 
for incident photon energies below 900~MeV and for $W$ below 1.6~GeV. These effects do not matter for 
analysis (1) of the asymmetry because they cancel since butanol and carbon data were measured under identical 
conditions.
 
The available data allow the helicity-dependent cross sections $\sigma_{1/2}$ and $\sigma_{3/2}$ to be extracted
in three different ways that have different systematic uncertainties. They can be computed as:
\begin{align}
\sigma_{1/2} &= \sigma_0(1+E)\nonumber \\
\sigma_{3/2} &= \sigma_0(1-E), 
\label{eq:Hel12}
\end{align}
where $E$ is the asymmetry measured in this experiment and $\sigma_0$ is the unpolarized cross section
measured with the liquid deuterium target \cite{Werthmueller_14}. For $E$ one can use the results from
the analysis version (1) or (2). We label the corresponding results for $E$ also with version (1) and
version (2). 

The third analysis, version (3), does not use the liquid deuterium data at all. It follows simply from:
\begin{align}
\sigma_{1/2} &= \frac{\sigma_{\rm sum}+\sigma_{\rm diff}}{2}\nonumber \\
\sigma_{3/2} &= \frac{\sigma_{\rm sum} -\sigma_{\rm diff}}{2}\, ,
\label{eq:Hel3}
\end{align}
with $\sigma_{\rm diff}$ and $\sigma_{\rm sum}$ as defined above. Ideally, all three analyses should give 
the same result within uncertainties. As shown in Sec.~\ref{sec:Res} this is in fact the case for the 
neutron data. For the proton data, again in the energy region of the $N(1535) 1/2^-$ resonance, 
versions (2) and (3) are affected by the detection efficiency problem and are discarded. 

\subsection{Systematic Uncertainties}
\label{sec:Sys}
The main systematic uncertainty of the $E$ asymmetry is related to the measurement of the beam and target
polarizations. The uncertainty of the photon polarization degree was determined to be $\pm2.7\%$ 
\cite{Tioukine_11}. The uncertainty of the target polarization was estimated as $\pm10\%$. This large uncertainty
is related to the fact that the polarization had to be renormalized to one measurement with a differently 
doped target. For the larger amount of data the polarization was varying across the target diameter 
in unpredictable ways. This means that the overall polarization of the target did not reflect the actual polarization in 
the area hit by the photon beam. In addition, for version (1) of the analysis of $E$ there is a small uncertainty
related to the subtraction of the carbon background (all other uncertainties e.g. from detection efficiencies cancel 
to a large extent in the ratio of Eq.~\ref{eq:E}). This uncertainty was estimated from the precision of the photon 
flux measurements and the determination of the target surface densities. It is on the order of 2.5\% and was added 
quadratically to the polarization degree uncertainties. The systematic uncertainties from this analysis for $E$,
and their propagation into the uncertainties of $\sigma_{1/2}$ and $\sigma_{3/2}$, are shown in the figures of 
Sec. \ref{sec:Res} as gray bands. The results from analysis version (2) are shown in the figures as an 
independent test of systematic effects. 

The overall normalization uncertainty of the unpolarized cross section from \cite{Werthmueller_14} also matters 
for the two helicity-dependent cross sections $\sigma_{1/2}$ and $\sigma_{3/2}$ (not for their
ratio) for the results from analyses versions (1) and (2). They are on the order of 7\% for production of
quasifree protons and on the order of 12\% for quasifree neutrons \cite{Werthmueller_14}. For version (3), the 
corresponding uncertainty stems from the overall normalization of the measurements with the butanol and carbon 
targets. These are of similar size except, as discussed above, for the reaction with quasifree protons in the
$N(1535) 1/2^-$ region.

The uncertainties quoted above were very conservatively estimated. There are further possibilities to check them
directly by the data. Significant contributions from the detection and identification of the $\eta$ mesons are
stringently limited by the fact that, as in \cite{Werthmueller_14}, no systematic discrepancies between the results
for the $\eta\rightarrow 2\gamma$ and $\eta\rightarrow 6\gamma$ decays modes were observed. A further check comes 
from the agreement between the different analysis versions, excluding the low-energy proton data, which are discussed 
in Sec.~\ref{sec:Res}. Finally, $\eta$ photoproduction in the threshold region has the property that almost 
exclusively the excitation of the $N(1535) 1/2^-$ contributes \cite{Krusche_15}. This means that in the threshold 
region the $E$ asymmetry should be unity and the relations $\sigma_{1/2}\approx 2\sigma_{0}$, 
$\sigma_{3/2}\approx 0$ should hold. For the free proton target, this behavior has been recently experimentally 
verified \cite{Senderovich_16} by the CLAS experiment and it can be used as a check of the absolute scale of 
the asymmetries.    

\section{Results}
\label{sec:Res}
As discussed in Sec.~\ref{sec:ExO}, the double polarization observable $E$ was extracted in two different ways. 
The difference of the two helicity dependent cross sections $\sigma_{1/2}$ and $\sigma_{3/2}$ was normalized 
to the carbon subtracted sum of them in analysis version (1). In analysis version (2) the 
normalization was done with the unpolarized cross section measured with a liquid deuterium target. 
The total asymmetries from analysis version (1) and also the helicity-dependent cross sections from this analysis
were summarized in a previous Letter \cite{Witthauer_16}. Here, we give a full account of the results
from all analyses including also the angular distribution of the asymmetries. In the first subsection, results are
shown as a function of the incident photon energy, i.e. these results are folded with Fermi motion. The results from
the kinematic reconstruction of the final state, which are not affected by Fermi motion, are discussed in the second
subsection. All results are statistically averaged over the $2\gamma$ and $6\gamma$ decays decay modes of the $\eta$ 
meson. 

\subsection{Results as a Function of the Incident Photon Energy}
\label{sec:ResEg}

\begin{figure}[t]
\centerline{
\resizebox{1.03\columnwidth}{!}{\includegraphics{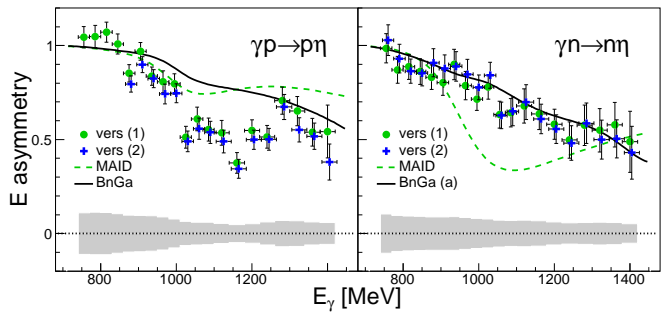}}}
\caption{Double-polarization observable $E$ as a function of the incident photon energy $E_{\gamma}$ 
for the proton (left) and the neutron (right). The experimental results are averaged over both decay 
channels $\eta\to2\gamma$ and $\eta\to6\gamma$. They are compared to Fermi-folded model results from
the BnGa \cite{Anisovich_15} and MAID \cite{Chiang_02} model. For better readability, the points 
from version (2) are shifted by $+5$ MeV with respect to version (1). The systematic uncertainties 
are indicated by the gray shaded areas.}
\label{fig:EEg}       
\end{figure}

\begin{figure*}[thb]
\centerline{
\resizebox{0.97\textwidth}{!}{\includegraphics{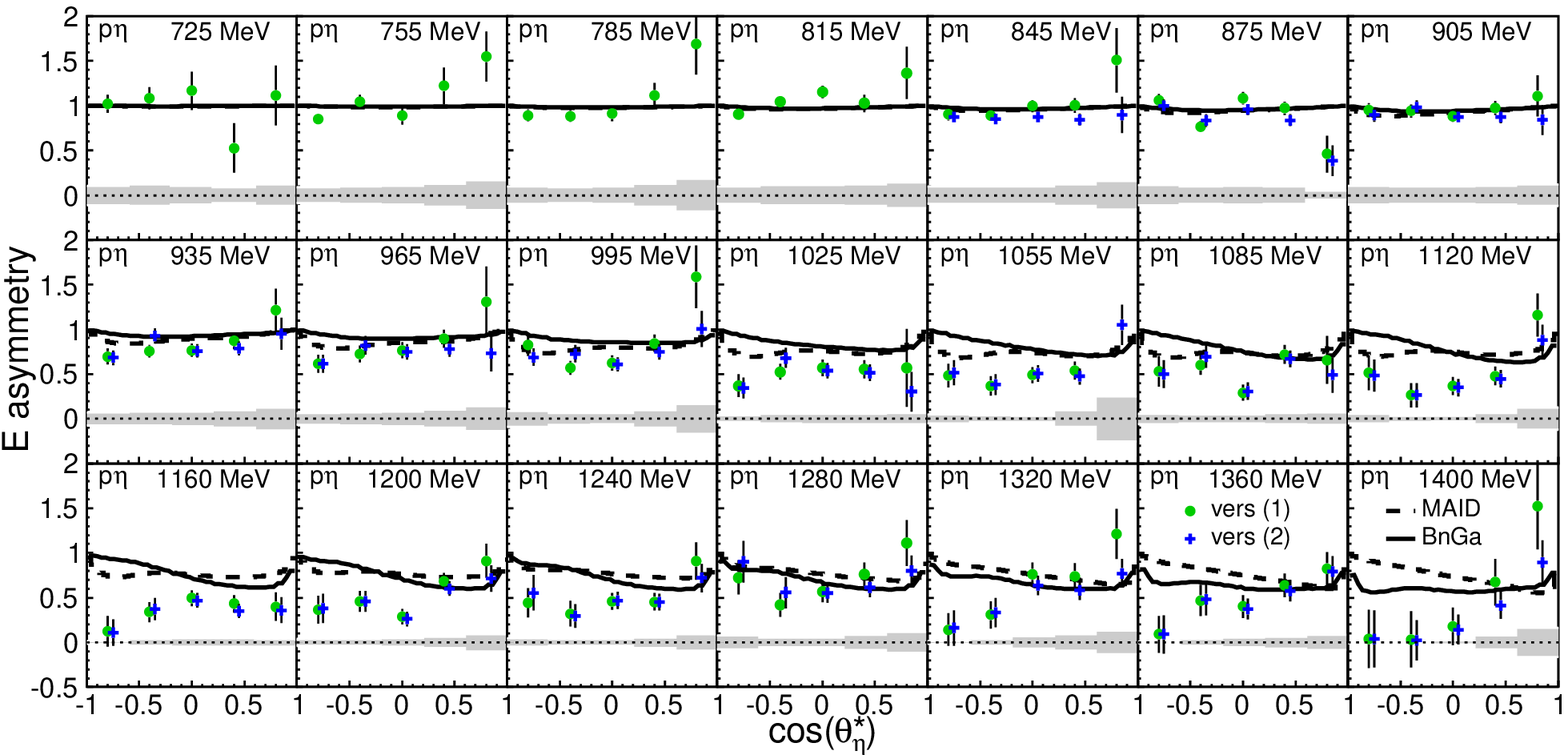}}}
\caption{(Color online) Angular distributions for the double-polarization observable $E$ for the recoil proton for 
bins of incident photon energy. For better visibility, the points of version (2) (blue crosses) were shifted 
by $\Delta \cos{(\theta_{\eta}^{\ast})}=+ 0.05$ with respect to version (1) (green dots). The systematic 
uncertainties are indicated by the gray shaded areas. The Fermi-folded model predictions by BnGa 
\cite{Anisovich_15} and MAID \cite{Chiang_02} are indicated as solid and dashed lines, respectively.}
\label{fig:dcsEEgp}       
\vspace{0.5cm}
\centerline{
\resizebox{0.97\textwidth}{!}{\includegraphics{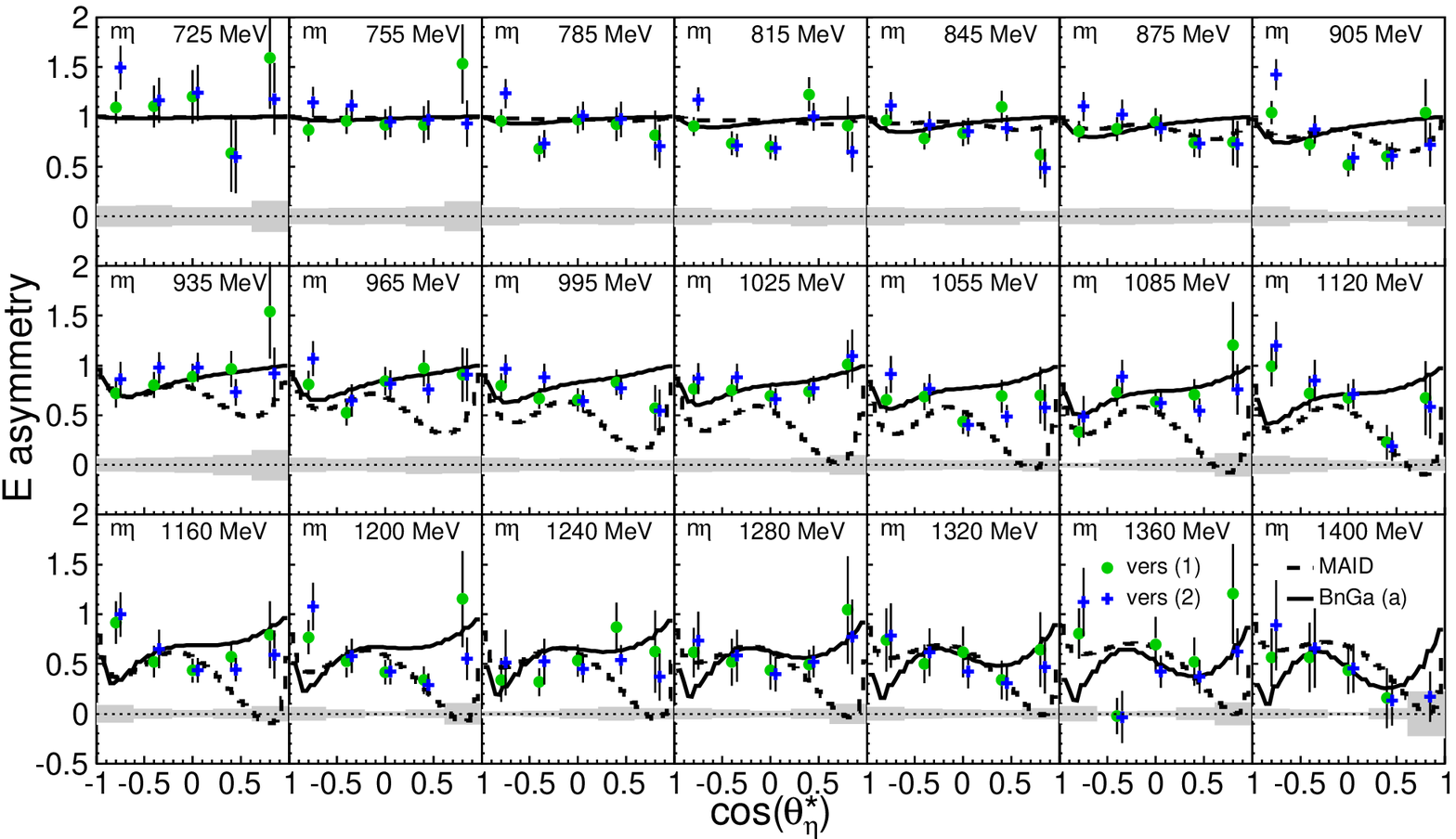}}}
\caption{(Color online) Angular distributions for the double-polarization observable $E$ for the recoil neutron for 
bins of incident photon energy. For better visibility, the points of version (2) (blue crosses) were shifted 
by $\Delta \cos{(\theta_{\eta}^{\ast})}=+ 0.05$ with respect to version (1) (green dots). The systematic 
uncertainties are indicated by the gray shaded areas. The Fermi-folded model predictions by BnGa 
\cite{Anisovich_15} (model with interference of the $N(1535)$ and the $N(1650)$) and MAID \cite{Chiang_02} 
are indicated as solid and dashed lines, respectively.}
\label{fig:dcsEEgn}       
\end{figure*}

The results for the two analysis versions as a function of the incident photon energy for quasifree
reactions of protons and neutrons is shown in Fig.~\ref{fig:EEg}. The angular distributions of this 
observable are summarized in Fig.~\ref{fig:dcsEEgp} for protons and in Fig.~\ref{fig:dcsEEgn} for neutrons. 
Apart from the low-energy region for the proton, the results from both analysis versions are shown together 
with the Fermi-folded model predictions from the MAID \cite{Chiang_02} and BnGa \cite{Anisovich_15} groups. 
The results from the analysis using the carbon background subtraction (version(1)), and from the analysis normalized 
to the measurement with the liquid deuterium target (version(2)), are in good agreement. As predicted by all models,
and also consistent with the experimental results for a free proton target \cite{Senderovich_16}, the 
$E$ asymmetry is consistent with unity within uncertainties from threshold throughout the $N(1535) 1/2^-$   
resonance region. At higher incident photon energies, for the proton as well as for the neutron target,
$E$ decreases, which indicates rising contributions from higher partial waves. However, $E$ does not become
much smaller than $+0.5$, which means that over the whole energy range $\sigma_{1/2}\gtrsim 3\sigma_{3/2}$, 
so that contributions from $J=1/2$ states are dominant. For the total
asymmetry, the predictions from both models \cite{Chiang_02,Anisovich_15} are similar for the proton and
disagree significantly with the experimental data in the energy range between 1.0 - 1.2~GeV. For the neutron,
the BnGa analysis is quite close to the data and the MAID prediction disagrees again for the energy range 
between 1.0 - 1.2~GeV, which can be traced to an unrealistically large contribution of the $N(1675) 5/2^-$
resonance. For the BnGa analysis, the results for the model based on an interference in the $S_{11}$ sector
are shown, but the other model versions give similar results. 

The angular distributions in Figs.~\ref{fig:dcsEEgp} and \ref{fig:dcsEEgn} show more details. They are 
of course flat in the threshold range where the $S_{11}$ wave dominates. At higher photon energies they 
develop more structure and can certainly help to constrain future partial-wave analyses. 
For such analyses, the results discussed in the next section for the kinematically reconstructed final state,
eliminating Fermi motion effects, are better suited. 

\subsection{Results as a Function of the Invariant Mass of the Final State}
\label{sec:ResW}
\begin{figure*}[bt]
\centerline{
\resizebox{0.90\textwidth}{!}{\includegraphics{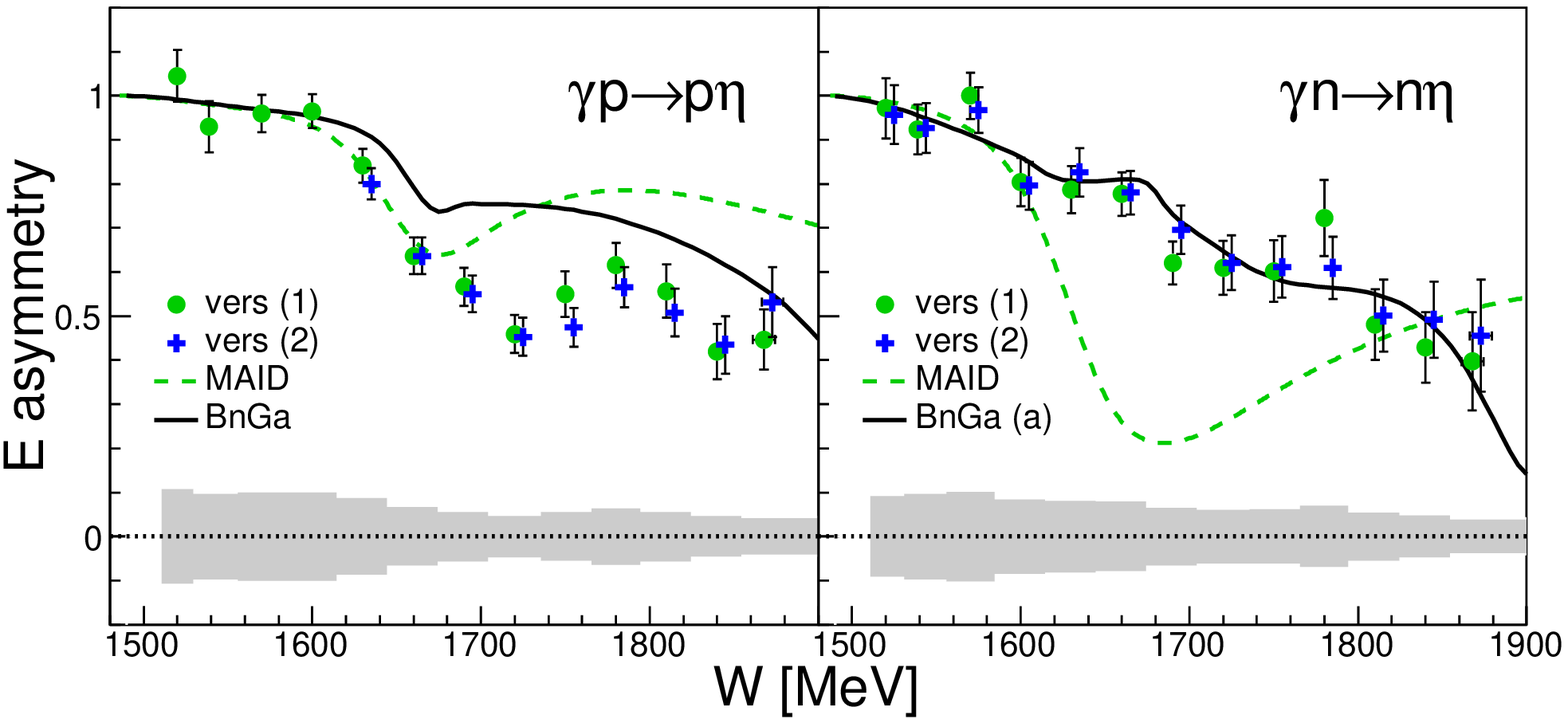}}}
\caption{(Color online) Double-polarization observable $E$ for the proton (left) and the neutron (right) 
shown as a function of the reconstructed c.m. energy. The results were averaged over both decay channels 
$\eta\to2\gamma$ and $\eta\to6\gamma$. The results are compared to model calculations by 
BnGa \cite{Anisovich_15} (neutron model with interference of the $N(1535)$ and the $N(1650)$) and 
MAID \cite{Chiang_02}. For better visibility, the points from version (2) were shifted by $+5$ MeV 
with respect to version (1). The systematic uncertainties for analysis (1) are indicated by the gray 
shaded areas.}
\label{fig:PolE}  
\vspace{0.5cm}     
\centerline{
\resizebox{0.99\columnwidth}{!}{\includegraphics{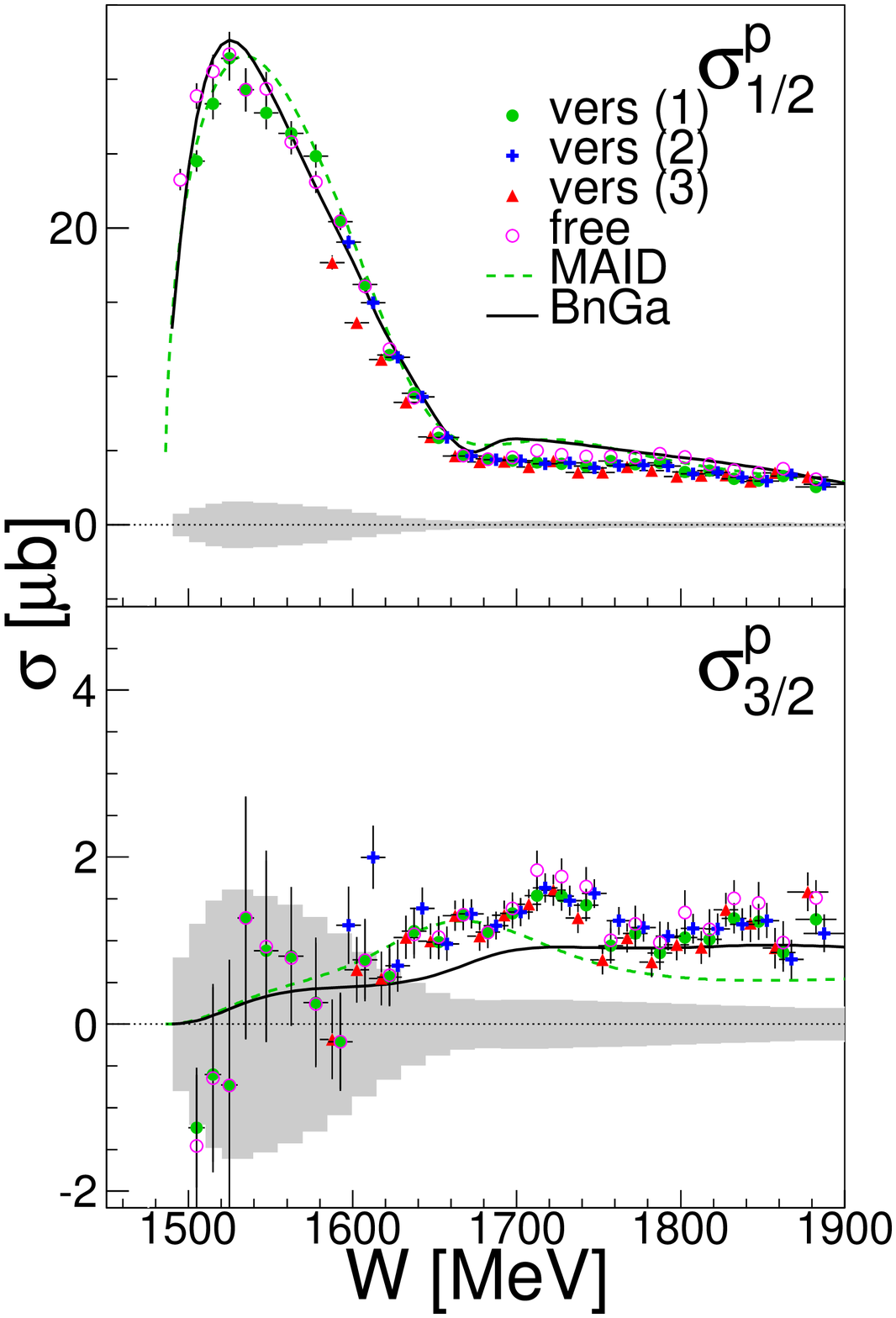}}
\resizebox{0.99\columnwidth}{!}{\includegraphics{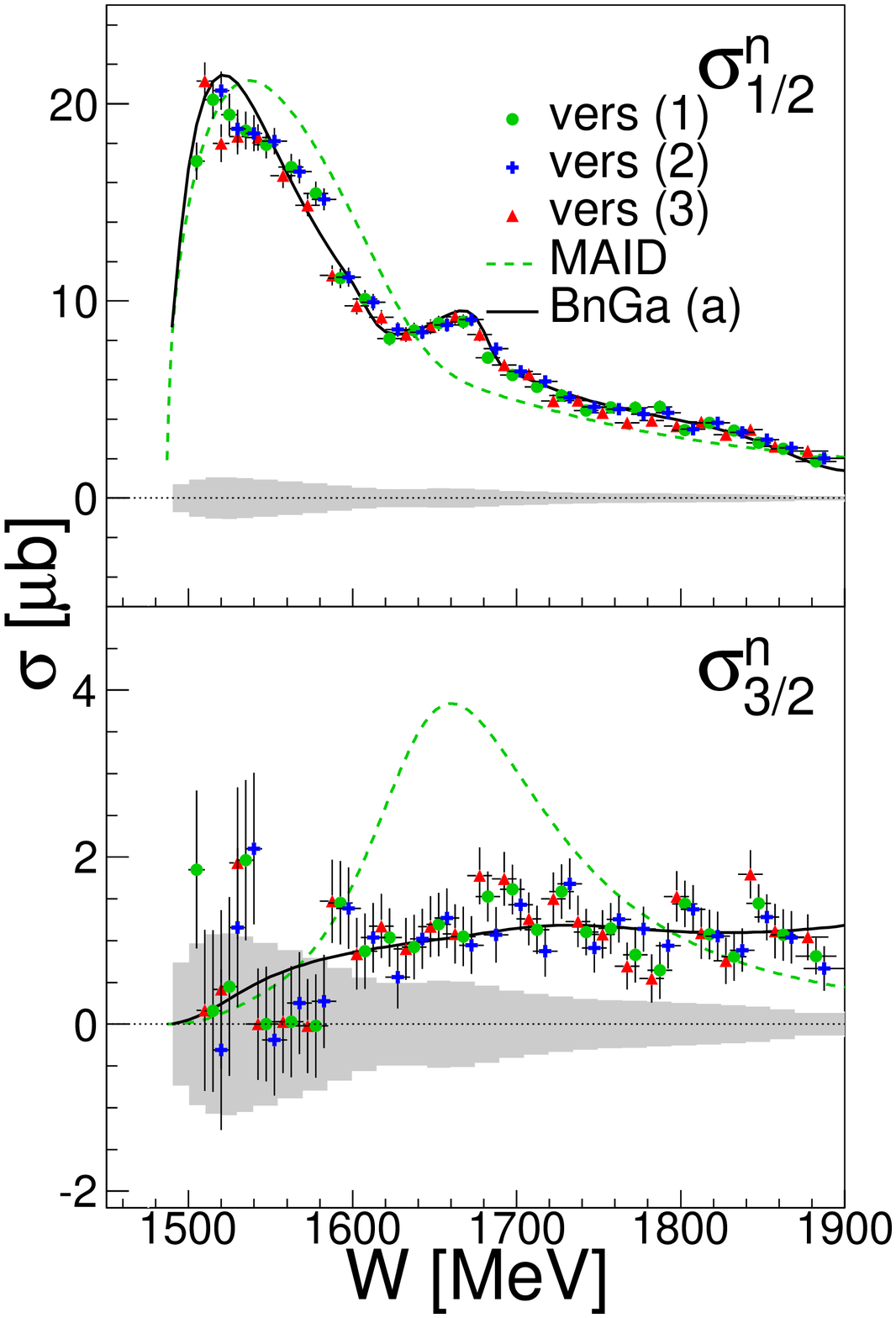}}
}
\caption{(Color online) Helicity-dependent cross sections $\sigma_{1/2}$ and $\sigma_{3/2}$  for the proton (left) 
and the neutron (right) as a function of the  reconstructed c.m. energy. The results were averaged over 
both decay channels $\eta\to2\gamma$ and $\eta\to6\gamma$ and  are compared to model calculations 
by BnGa \cite{Anisovich_15} (neutron model with interference of the $N(1535)$ and the $N(1650)$) and 
MAID \cite{Chiang_02}. For better visibility, the points from version (2) and version (3)  
were shifted by $\pm5$ MeV with respect to version (1). The systematic uncertainties for analysis (1) 
are indicated by the gray shaded areas. For the proton, results are also shown (labeled `free') when for version (1)
of the analysis the unpolarized cross section $\sigma_0$ is taken from free-proton data \cite{McNicoll_10}.}
\label{fig:Tot}       
\end{figure*}

The results for the double-polarization observable $E$ as a function of the reconstructed c.m. energy $W$ 
are shown in Fig.~\ref{fig:PolE} for the proton (left) and neutron (right). The general behavior is 
similar to the results as function of incident photon energy, but due to the better energy resolution 
achieved after removal of Fermi smearing there is a small peak-like structure visible for the reaction 
off neutrons at $W$ around 1680~MeV. Again, apart from the low-energy region for the proton target the
results from carbon subtraction, analysis (1), and from deuterium normalization, analysis (2), are in good 
agreement. 

The data are compared to the model predictions from the BnGa \cite{Anisovich_15} and MAID 
\cite{Chiang_02} model analyses. All models reproduce the unity value of the asymmetry at threshold, but 
for the proton target, agreement is surprisingly poor at higher energies. The BnGa model overestimates 
the asymmetry above $W\approx1.6$~GeV, the MAID model above 1.65~GeV. It seems that in particular around 
1.7~GeV some components with higher spin are still missing in both models. For the neutron, the BnGa model 
version (a) \cite{Anisovich_15} reproduces the data quite well. This is not surprising because this model 
was fitted to reproduce the data for the unpolarized cross section from Ref.~\cite{Werthmueller_14} with 
a tuning of the interference pattern in the $S_{11}$ sector. Consequently, it reproduces the bump-like 
structure around 1680~MeV with contributions from the $\sigma_{1/2}$ component to the total cross section. 
The width of the structure in $\sigma_{1/2}$ is approximately 30 MeV (FWHM), which is comparable to the 
experimental resolution in that energy range. This was taken into account for the BnGa fits. The model 
results were folded with the experimental resolution before they were compared to the data.
The result from the MAID model disagrees completely, because there the cross section access in this energy 
range stems from the $N(1675) 5/2^-$ state, which pushes the asymmetry in the opposite direction.   

Using Eqs.~\ref{eq:Hel12},\ref{eq:Hel3} one can now extract the helicity-dependent cross sections
$\sigma_{1/2}$ and $\sigma_{3/2}$ in the three different ways discussed in Sec.~\ref{sec:ExO}.
The results are shown in Fig.~\ref{fig:Tot}. The analysis (version (1)) with the smallest systematic 
uncertainties uses Eq.~\ref{eq:Hel12} with $E$ determined from the carbon subtraction method and 
combines it with the precise values of the unpolarized cross section $\sigma_{0}$ from \cite{Werthmueller_14}.
The systematic uncertainties shown in Fig.~\ref{fig:Tot} correspond to this analysis. However, apart from
the low-energy region for the proton the results from all three analyses are in good agreement. These 
results are of course statistically not independent and therefore should not be averaged. For example, for
analysis (1) and (2) in both cases identical values enter in the numerator $\sigma_{1/2}-\sigma_{3/2}$ 
for $E$ and identical values are used for $\sigma_{0}$. They are only limiting systematic uncertainties. 

\begin{figure*}[p]
\centerline{
\resizebox{0.98\textwidth}{!}{\includegraphics{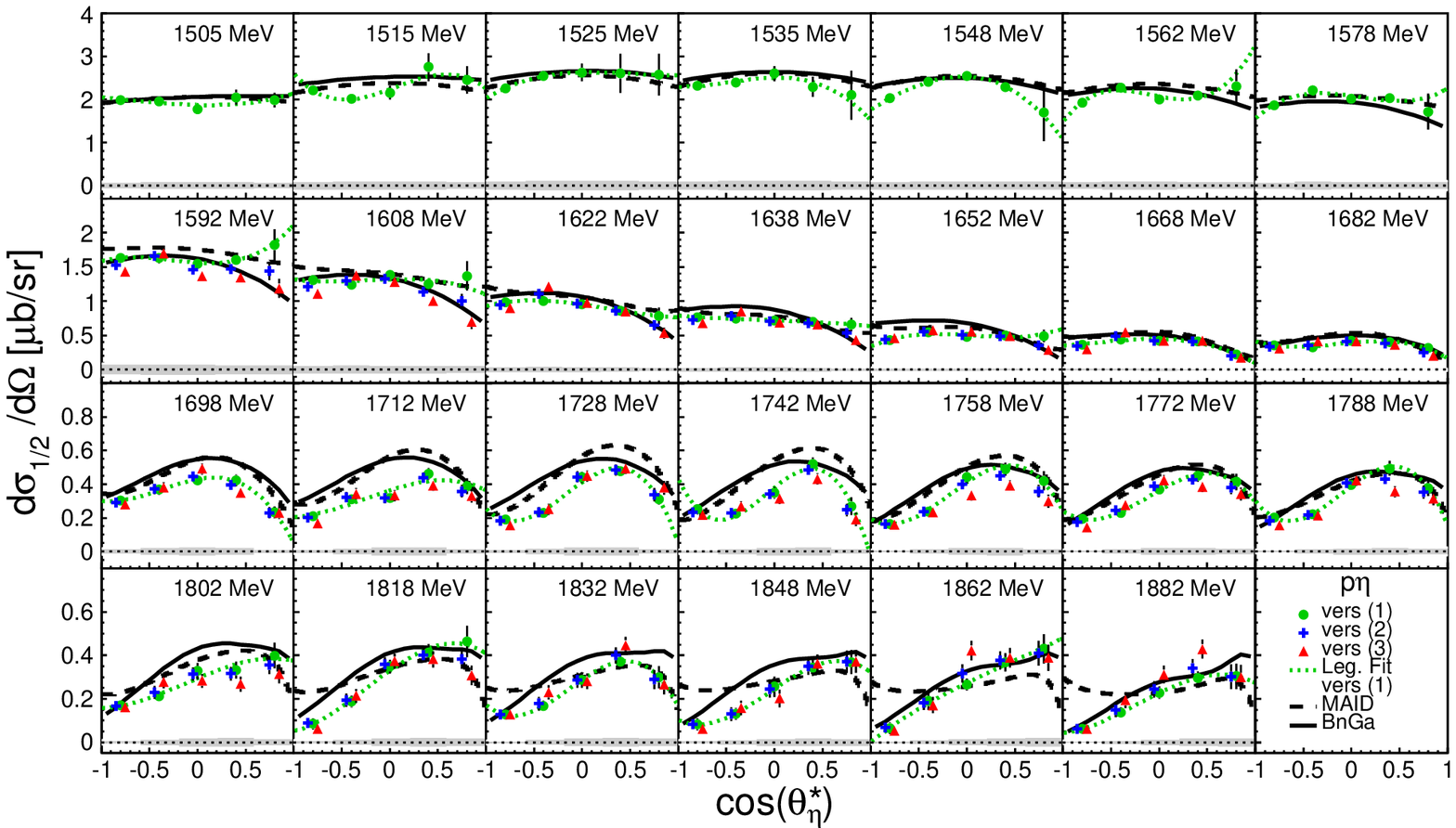}}}
\caption{(Color online) Angular distributions for the helicity-dependent cross section $\sigma_{1/2}$ for the 
proton. The results are shown in the c.m. frame of the $\eta$ meson and the final-state nucleon. For better 
visibility, the points of version (2) (blue crosses) were shifted by 
$\Delta \cos{(\theta_{\eta}^{\ast})}=+ 0.05$ with respect to version (1) (green dots). The systematic 
uncertainties are indicated by the gray shaded areas. The model predictions by BnGa \cite{Anisovich_15} 
and MAID \cite{Chiang_02} are indicated as solid and dashed lines, respectively.}
\label{fig:dcs12p}       
\vspace{0.5cm}
\centerline{
\resizebox{0.98\textwidth}{!}{\includegraphics{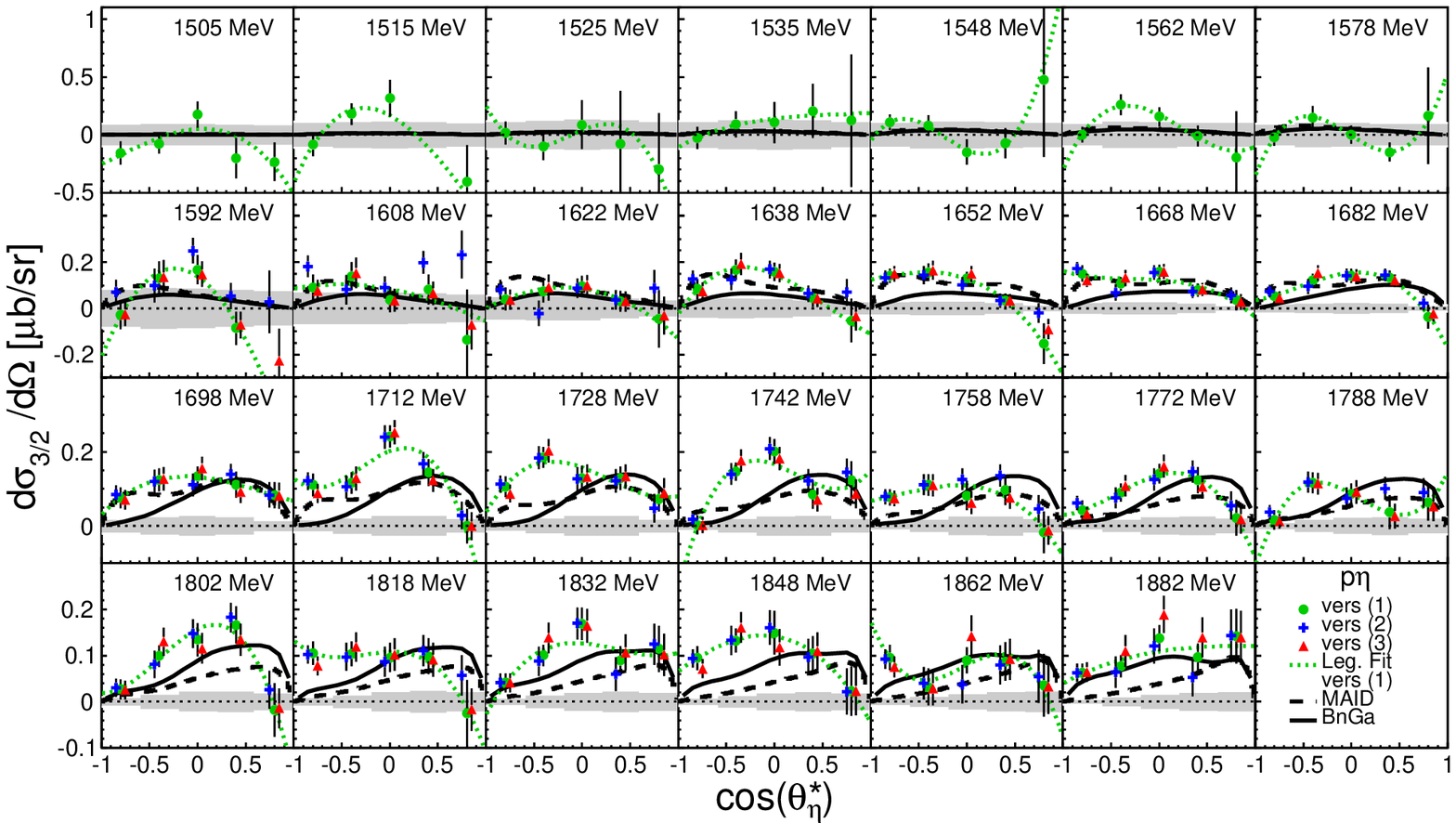}}}
\caption{(Color online) Angular distributions for the helicity-dependent cross section $\sigma_{3/2}$ for the 
proton. The results are shown in the c.m. frame of the $\eta$ meson and the final-state nucleon. For better 
visibility, the points of version (2) (blue crosses) were shifted by 
$\Delta \cos{(\theta_{\eta}^{\ast})}=+ 0.05$ with respect to version (1) (green dots). The systematic 
uncertainties are indicated by the gray shaded areas. The model predictions by BnGa \cite{Anisovich_15} 
and MAID \cite{Chiang_02} are indicated as solid and dashed lines, respectively.}
\label{fig:dcs32p} 
\end{figure*}

\begin{figure*}[p]
\centerline{
\resizebox{0.98\textwidth}{!}{\includegraphics{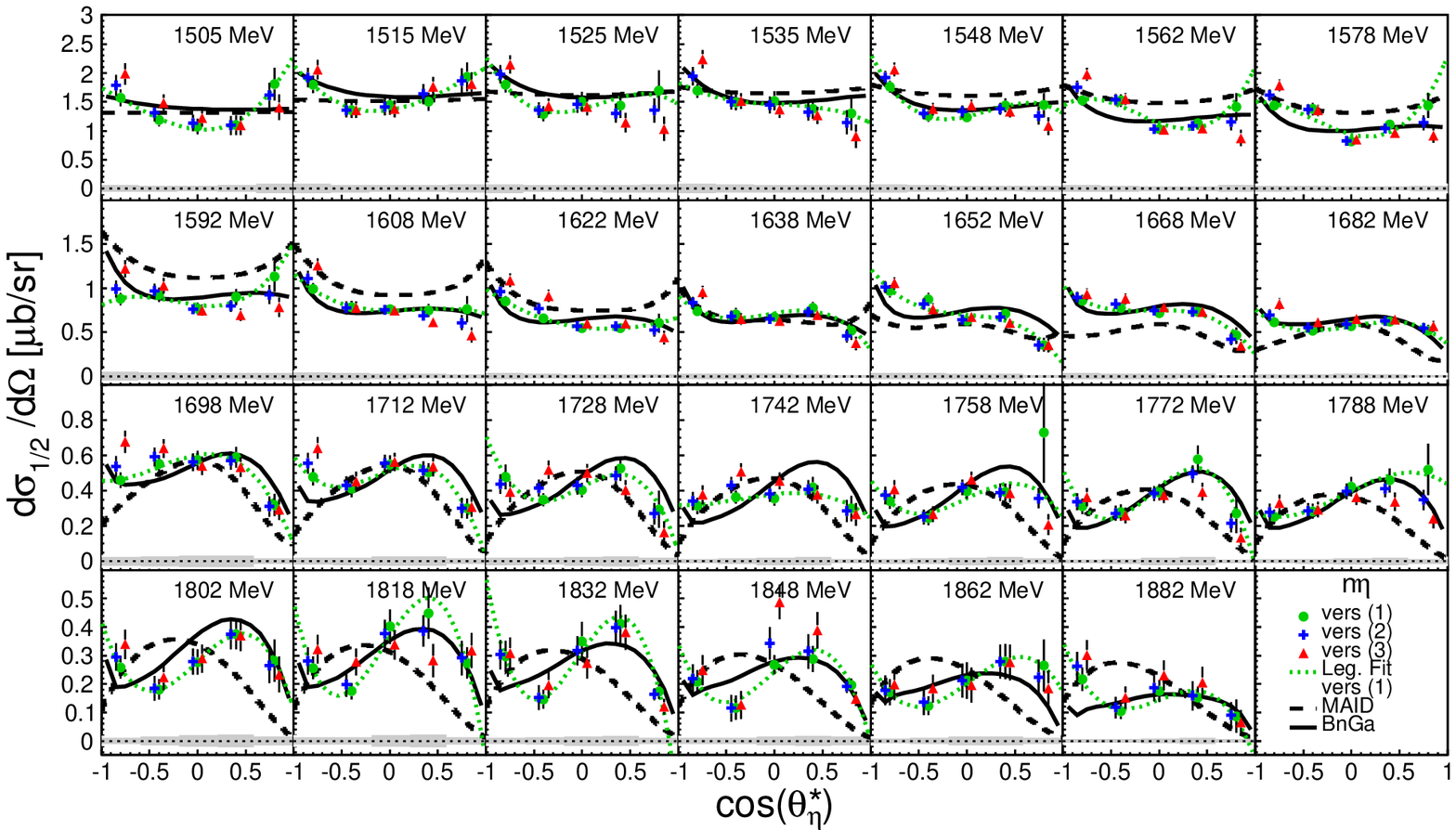}}}
\caption{(Color online) Angular distributions for the helicity-dependent cross section $\sigma_{1/2}$ for the 
neutron. The results are shown in the c.m. frame of the $\eta$ meson and the final-state nucleon. For better 
visibility, the points of version (2) (blue crosses) were shifted by 
$\Delta \cos{(\theta_{\eta}^{\ast})}=+ 0.05$ with respect to version (1) (green dots). The systematic 
uncertainties are indicated by the gray shaded areas. The model predictions by BnGa \cite{Anisovich_15} and 
MAID \cite{Chiang_02} are indicated as solid and dashed lines, respectively.}
\label{fig:dcs12n}       
\vspace{0.5cm}
\centerline{
\resizebox{0.98\textwidth}{!}{\includegraphics{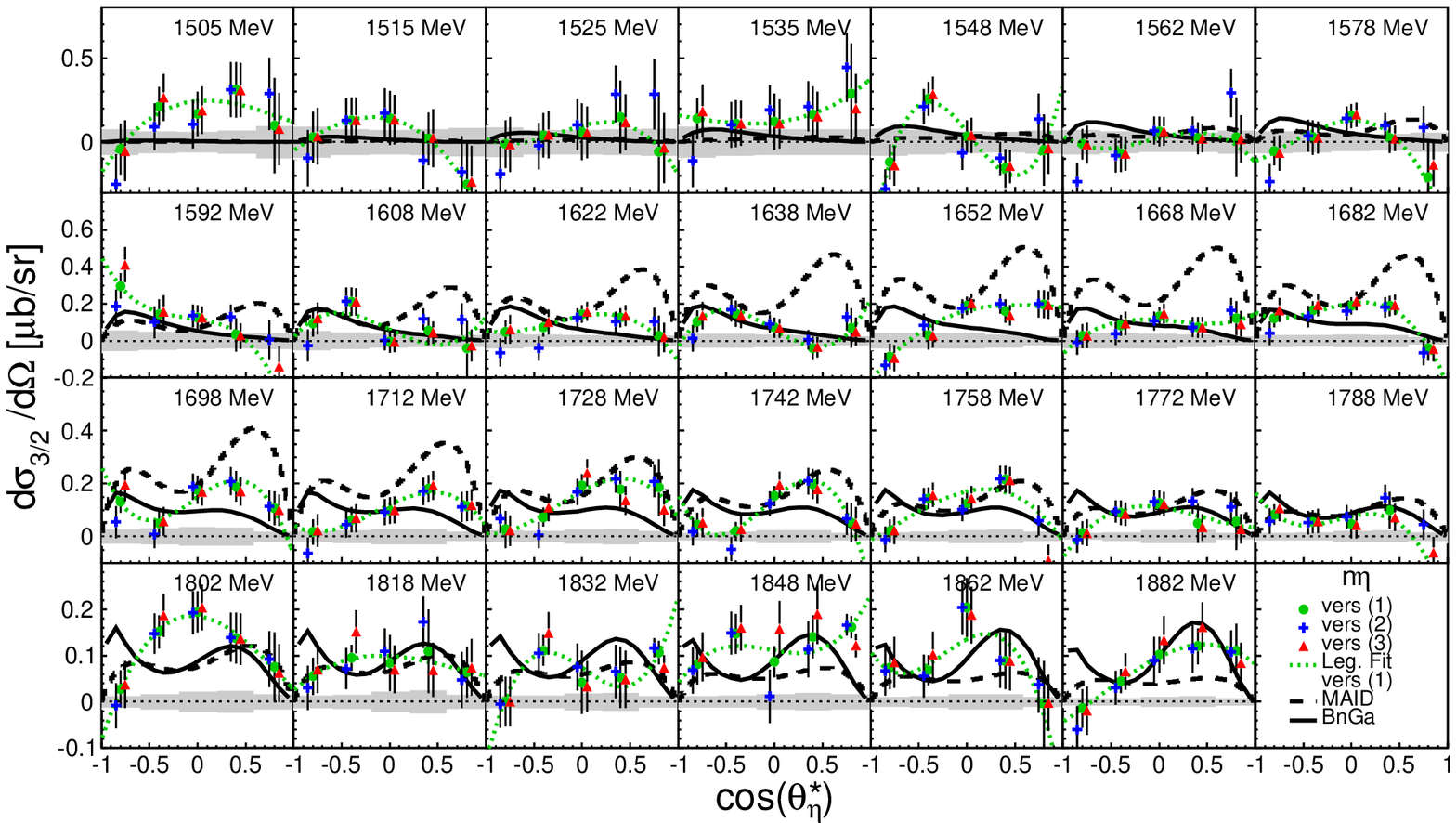}}}
\caption{(Color online) Angular distributions for the helicity-dependent cross section $\sigma_{3/2}$ for the 
neutron. The results are shown in the c.m. frame of the $\eta$ meson and the final-state nucleon. For better 
visibility, the points of version (2) (blue crosses) were shifted by 
$\Delta \cos{(\theta_{\eta}^{\ast})}=+ 0.05$ with respect to version (1) (green dots). The systematic 
uncertainties are indicated by the gray shaded areas. The model predictions by BnGa \cite{Anisovich_15} 
and MAID \cite{Chiang_02} are indicated as solid and dashed lines, respectively.}
\label{fig:dcs32n} 
\end{figure*}

\begin{figure*}[bth]
\centerline{
\resizebox{0.97\textwidth}{!}{\includegraphics{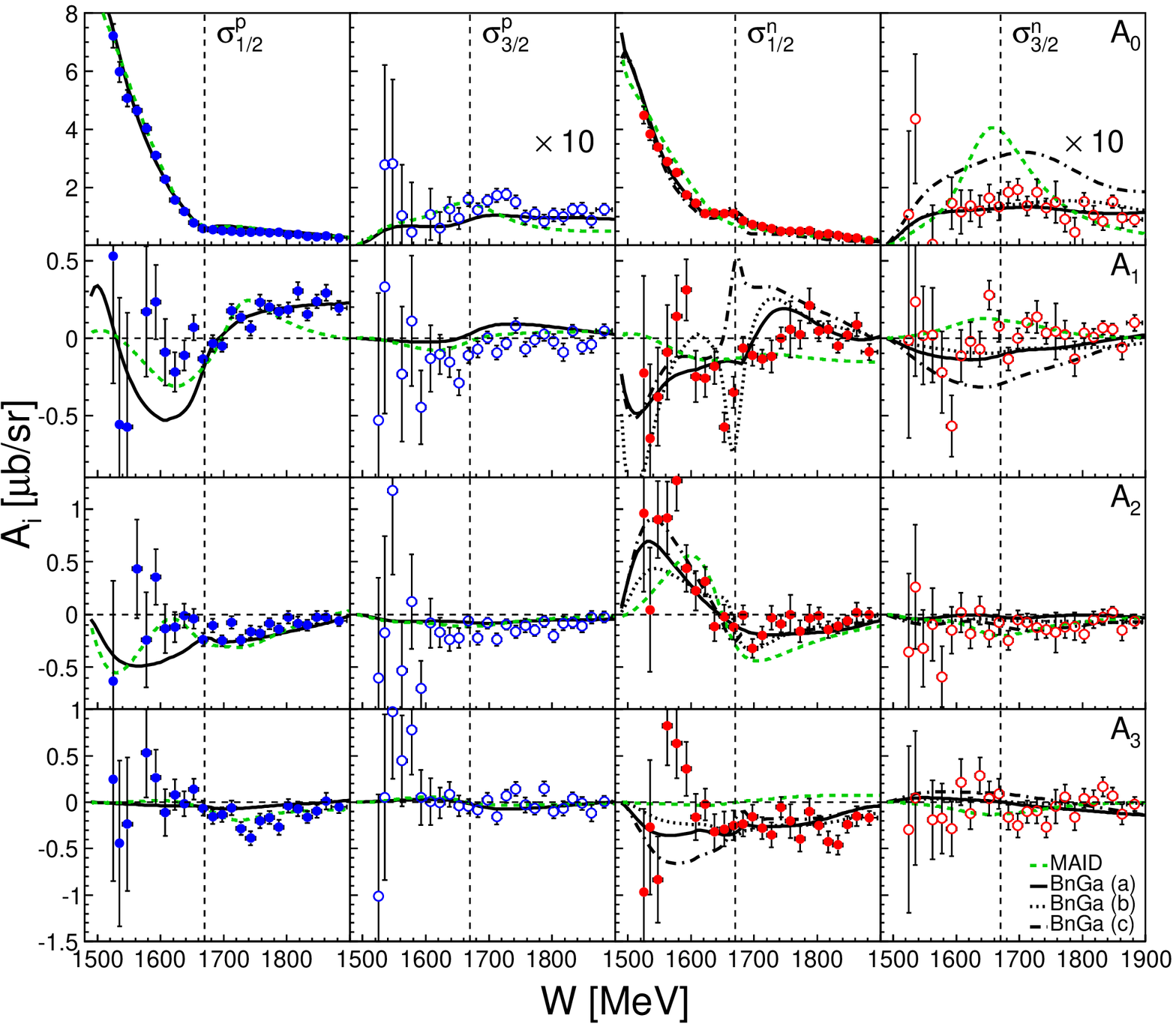}}}
\caption{(Color online) Legendre coefficients $A_0$ -- $A_{3}$ (rows) as defined in Eq.\ \ref{eq:LegPol}, which were
extracted from version (1). First column: coefficients for the helicity-$1/2$ state (solid circles) for 
the reaction on the proton. Second column: coefficients for the helicity-$3/2$ state (open circles) 
for the reaction on the proton. Third and fourth column: same for the reaction on the neutron. 
The experimental results (blue and red markers) are compared to the coefficients extracted from model 
predictions by MAID \cite{Chiang_02} (dashed green line) and BnGa \cite{Anisovich_15}. Three different BnGa 
models predictions are shown for the neutron. BnGa (b): fit with a narrow $N(1685)$ resonance with 
positive $A_{1/2}$ coupling (dotted line). BnGa (c): fit with a narrow $N(1685)$ resonance with 
negative $A_{1/2}$ coupling (dash-dotted line). BnGa (a): fit without a narrow resonance (solid line). 
The position of the narrow structure at $W=1685$ MeV in the neutron cross section is indicated by a 
dashed vertical line.}
\label{fig:Leg}       
\end{figure*}

Some interesting features of the data in Fig.~\ref{fig:Tot} can be discussed even without any 
results from reaction models. For the whole energy range the $\sigma_{3/2}$ part of the reaction is smaller 
than $\sigma_{1/2}$, underlining the importance of contributions from nucleon resonances with spin $J=1/2$. 

A very prominent feature for the neutron target is the narrow structure around $W=1.68$~GeV, 
which has no counterpart in $\sigma_{3/2}$. The cross section excess above the smoothly varying `background' 
is on the order of 2 $\mu b$ for $\sigma_{1/2}$, while the $\sigma_{3/2}$ cross section in this energy range 
is on an absolute scale of only 1 $\mu b$ and structureless. The structure previously observed in the 
unpolarized cross section is therefore clearly related to the helicity-1/2 part of the reaction. 
Nucleon resonances with spin larger than $J=1/2$ can also contribute to $\sigma_{1/2}$, but in most known 
cases they contribute stronger to $\sigma_{3/2}$ and there are no known examples where a spin $J\geq 3/2$ 
state contributes exclusively to $\sigma_{1/2}$ \cite{PDG_16}. This makes it very unlikely that the 
narrow structure is related to nucleon resonances with spin $J>1/2$. 

Shown in Fig.~\ref{fig:Tot} are also the model predictions from BnGa \cite{Anisovich_15} and MAID 
\cite{Chiang_02}. For the BnGa neutron model the version with a fine-tuned interference in the $S_{11}$ sector 
is shown, but the other versions are not much different. They agree quite well with the data. The results
from the MAID model have the known problem with the contribution from the $N(1675) 5/2^-$ state.

The BnGa results do not describe the proton data well above $W=1.65$~GeV. They agree 
of course with the unpolarized cross section from McNicoll et al. \cite{McNicoll_10}, because they
have been fitted to it, but not so good with the split into $\sigma_{1/2}$ and $\sigma_{3/2}$ contributions
suggested by the data. This disagreement does not disappear when instead of the quasi-free proton cross section
given in \cite{Werthmueller_13,Werthmueller_14} the free proton cross section from \cite{McNicoll_10} is used as
$\sigma_0$ in Eq.~\ref{eq:Hel12} (results shown as open magenta circles at the left-hand side of Fig.~\ref{fig:Tot}).  

In the total $\gamma p\rightarrow p\eta$ cross section \cite{McNicoll_10} there is 
a small, narrow dip exactly at the same $W$ where the neutron cross section shows the narrow bump.
This could have been a hint that in fact the neutron bump and proton dip could be related 
due to an interference that is constructive for the neutron and destructive for the proton. 
The present $\sigma_{1/2}$ data do not show any dip-like structure around $W\approx1.68$~MeV,
they are flat in this range. Instead, the $\sigma_{3/2}$ data show a little bump at slightly 
higher energy ($W\approx1.72$~GeV) and then the (unpolarized) sum of these two excitation functions has
an effective little dip-like structure around 1.68~GeV. 

The small bump in $\sigma_{3/2}$ could be due to
a contribution from the $N(1720)3/2^{+}$ state, but certainly more refined partial-wave analyses are 
necessary to confirm this. This structure is not visible for the neutron, but in that case simply the
statistical quality of the data may be insufficient. Independently on the nature of this structure,
the fact that it appears in $\sigma_{3/2}$ makes it much less probable that the bump in the neutron
excitation function and the dip in the proton excitation function are related phenomena. This problem is
also apparent in the comparison of the data to the model predictions. Both models fail to reproduce
the little peak in the $\sigma_{3/2}$ part of the cross section but rather shift this structure to the
$\sigma_{1/2}$ part. 
  
The angular distributions of the helicity-dependent cross sections are shown in 
Figs.~\ref{fig:dcs12p} and \ref{fig:dcs32p} for the proton and in Figs.~\ref{fig:dcs12n} and \ref{fig:dcs32n}
for the neutron together with the BnGa \cite{Anisovich_15} and MAID \cite{Chiang_02} model predictions.
It is obvious that, especially at higher energies, the new data will have significant impact when they
are included into the fits. Also shown, for a phenomenological analysis, are the results of fits of 
the present data with Legendre polynomials up to third order using: 
\begin{equation}
\frac{d\sigma}{d\Omega}(W,\mbox{cos}(\theta_{\eta}^{\star})) = \frac{q_{\eta}^{*}(W)}{k_{\gamma}^{*}(W)}
\sum_{i=0}^{3} A_i(W) P_i(\mbox{cos}(\theta_{\eta}^{\star}))\,,
\label{eq:LegPol}
\end{equation}
where $q_{\eta}^{*}$ and $k_{\gamma}^{*}$ are the $\eta$ and photon momenta in the center-of-mass 
frame, respectively, and  $A_i(W)$ are the Legendre coefficients. The fit results for analysis version (1)
are shown in Figs.~\ref{fig:dcs12p}-\ref{fig:dcs32n} as dotted (green) lines. 

The Legendre coefficients extracted from these fits are shown in Fig.~\ref{fig:Leg}. In order to keep 
the figure readable only the results from analysis version (1) are shown as data points with error bars 
(the results from the other analyses do not differ in any relevant aspect). Also shown are the Legendre
coefficients for the predictions of the MAID \cite{Chiang_02} and BnGa \cite{Anisovich_15} models, extracted
with the same fitting procedure using Eq.~\ref{eq:LegPol}. For the latter, for the neutron target, 
all three different solutions from \cite{Anisovich_15} are shown. These are BnGa (a), for which the
bump in the neutron excitation function around 1.68~GeV is reproduced by a fine tuning of interferences
in the S$_{11}$ sector, BnGa (b) where a narrow $P_{11}$ resonance with positive interference term to the
leading $S_{11}$ partial wave is introduced, and BnGa (c) where such a resonance with negative 
interference term contributes. The most sensitive observable to discriminate between these different model
approaches is the $A_1$ coefficient of the neutron $\sigma_{1/2}$ data. This is so, because an interference
between a $S_{11}$ and a $P_{11}$ wave introduces a cos$(\theta^{\star})$ term into the angular 
distributions, which is reflected in the $A_{1}$ coefficient, while an $S_{11}$ - $S_{11}$ interference 
results in flat angular distributions. The comparison of data and model results in Fig.~\ref{fig:Leg}
clearly rules out the case of a $S_{11}$ - $P_{11}$ interference with negative sign (dash-dotted
black line). However, the solution of a narrow $P_{11}$ state in interference with the $S_{11}$ wave 
with a positive sign (dotted line) is even closer to the data than the $S_{11}$ - $S_{11}$ interference 
(solid line).  

\section{Summary and Conclusions}
\label{sec:Sum}
In summary, precise results for the helicity decomposition of the cross sections of the reactions 
$\gamma p\rightarrow p\eta$ and $\gamma n\rightarrow n\eta$ measured with quasifree nucleons bound in the
deuteron have been obtained. These data confirm many previously known aspects of $\eta$ photoproduction and
add key information to the interpretation in particular of the narrow structures seen in their
excitation functions around invariant masses of $W\approx1.68$ GeV. The most important one is that
the narrow structure previously observed in the total cross section of the $\gamma n\rightarrow n\eta$
reaction appears only in the $\sigma_{1/2}$ part of the cross section and is thus almost certainly related
to the $S_{11}$ and/or $P_{11}$ partial waves. At the same time, the data with coincident protons show
that the small dip observed in the total cross section of $\eta$ production from free protons at a similar
energy can be assigned to structure in the $\sigma_{3/2}$ part of the reaction so that it is unlikely that 
both phenomena have the same cause. Finally, a comparison of the angular distributions, in particular
the coefficient $A_1$ of their Legendre expansion, to model predictions gives some preference to an
interference between the dominating $S_{11}$ wave with a narrow $P_{11}$ state. However, these results
are statistically not very significant. Obviously, final conclusions from these new data can only be drawn 
after much more detailed model analyses, which are underway.

\begin{acknowledgments}
We wish to acknowledge the outstanding support of the accelerator group and operators of MAMI. 
This work was supported by Schweizerischer Nationalfonds (200020-156983, 132799, 121781, 117601), 
Deutsche For\-schungs\-ge\-mein\-schaft (SFB 443, SFB 1044, SFB/TR16), the INFN-Italy, 
the European Community-Research Infrastructure Activity under FP7 programme (Hadron Physics, 
grant agreement No. 227431), 
the UK Science and Technology Facilities Council (ST/J000175/1, ST/G008604/1, ST/G008582/1,ST/J00006X/1, and 
ST/L00478X/1), 
the Natural Sciences and Engineering Research Council (NSERC, FRN: SAPPJ-2015-00023), Canada. This material 
is based upon work also supported by the U.S. Department of Energy, Office of Science, Office of Nuclear 
Physics Research Division, under Award Numbers DE-FG02-99-ER41110, DE-FG02-88ER40415, and DE-FG02-01-ER41194 
and by the National Science Foundation, under Grant Nos. PHY-1039130 and IIA-1358175.
\end{acknowledgments}

\end{document}